\documentclass[useamsfonts]{pasj01}
\usepackage{natbib}
\usepackage{aas_macros}
\usepackage{url}
\usepackage{bm}
\usepackage{lineno}
\usepackage{multirow}
\usepackage{tabularx}
\usepackage{booktabs}

\newcommand{\simgt}{\lower.5ex\hbox{$\; \buildrel > \over \sim \;$}}
\newcommand{\simlt}{\lower.5ex\hbox{$\; \buildrel < \over \sim \;$}}

\begin{document}
\SetRunningHead{T. Hamana}{Mitigating an excess up-scattering mass bias on
  the weak lensing cluster mass estimates}
\Received{2022/8/11}
\Accepted{2022/10/13}
\Published{}

\title{An empirical method for mitigating an excess up-scattering mass bias
  on the weak lensing mass estimates for shear-selected cluster samples}

%
%%% begin:list of authors
%
% Do NOT capitalize all letters in "textsc".
\author{Takashi \textsc{Hamana}\altaffilmark{1}}
\altaffiltext{1}{National Astronomical Observatory of Japan, Mitaka,
  Tokyo 181-8588, Japan}
%%%
%
%%% end:list of authors
%
%\email{hamana.tk@nao.ac.jp}

%% `\KeyWords{}' always has to be placed before `\maketitle'.

\maketitle

\begin{abstract}
An excess up-scattering mass bias on a weak lensing
cluster mass estimate is a statistical bias that an observed
weak lensing mass ($M_{\rm obs}$) of a cluster of galaxies is, in a
statistical sense, larger than
its true mass ($M_{\rm true}$) because of a higher chance of
up-scattering than that of down-scattering due to random noises in a weak
lensing cluster shear profile.
This non-symmetric scattering probability is caused by a monotonically
decreasing cluster mass function with increasing mass.
We examine this bias (defined by $b=M_{\rm obs}/M_{\rm true}$) in weak
lensing shear-selected clusters, and 
present an empirical method for mitigating it.
In so doing,
we perform the standard weak lensing mass estimate of realistic mock
clusters,
and find that the weak lensing mass estimate based on the standard $\chi^2$
analysis gives a statistically correct confidence intervals, but
resulting best-fitting masses are biased high on average. 
Our correction method uses the framework of the standard Bayesian
statistics with the prior of the
probability distribution of the cluster mass and concentration parameter
from recent empirical models.
We test our correction method using mock weak lensing clusters, and find
that the method works well with resulting corrected $M_{\rm obs}$-bin averaged mass biases
being close to unity within $\sim 10$ percent.
We applied the correction method to weak lensing shear-selected cluster sample of 
\citet{2020PASJ...72...78H},
and present bias-corrected weak lensing cluster masses.
\end{abstract}

%\linenumbers

%
%%%%%% Introduction %%%%%%%%%%%%%%%%%%%%%%%%%%%%%%%%%%%%%%%%%%%%%%%%%%
%
\section{Introduction}
\label{sec:intro}
Clusters of galaxies have been played important roles in modern
cosmology and studies of structure formation: Cluster abundance and its
time evolution have been used as a unique cosmological probe
\citep{2011ARA&A..49..409A}, and clusters are valuable sites to study
physical processes of hierarchical structure formation
\citep[see][for a review]{2012ARA&A..50..353K}. 
Total mass of clusters of galaxies, which consists of dark matter and
baryonic components, is one of its most fundamental properties to link
observation and theory.
Therefore an accurate estimate of a cluster mass is of essential
importance to cluster science.
Several methods have been used to estimate
individual cluster masses, including the galaxy kinematics, X-ray or/and
SZ observations, and weak lensing observations \citep[see][for a
  comprehensive review]{2019SSRv..215...25P}.
Among them, the weak lensing method is unique in the sense that it does
not rely on assumptions about the dynamical status of cluster galaxies or
the hydro-dynamical status of intra-cluster gas and underlying gravitational
potential.

Every method for estimating cluster mass has its own
scatter and bias, and a weak lensing cluster mass estimate is no
exception.
It is known to be affected by some noises and uncertainties in
measurements and modeling 
\citep[see comprehensive reviews by][and references
  therein]{2019SSRv..215...25P,2020A&ARv..28....7U}; including, among
others, 
(1) noises from intrinsic galaxy shapes,
(2) a projection effect by un-associated
structures in the same line-of-sight of a cluster,
(3) deviations from an assumed cluster mass distribution
(most commonly, a spherical Navvarro-Frenk-White (NFW) model;
\citet{1997ApJ...490..493N}),
(4) an uncertainty in the observational determination of a cluster center,
and (5) an uncertainty in the redshift distribution of source galaxies
inferred from photometric redshift information,
Some, for example the above (1) and (2), are unavoidable, and thus need
to be properly taken into account in a covariance of a statistical analysis
of cluster mass estimate.
Others require a better understanding of their influence on a cluster
mass estimate, and improved techniques need to be developed to reduce
their impacts. 
In this paper, we focus on another bias arising in estimating cluster
mass with commonly used statistical analysis of noisy weak lensing data,
which we will describe in detail below.

%
%%%%% Fig demo_exupscatbias.ps --- fig-1
%
\begin{figure}
\begin{center}
  \includegraphics[height=82mm,angle=270]{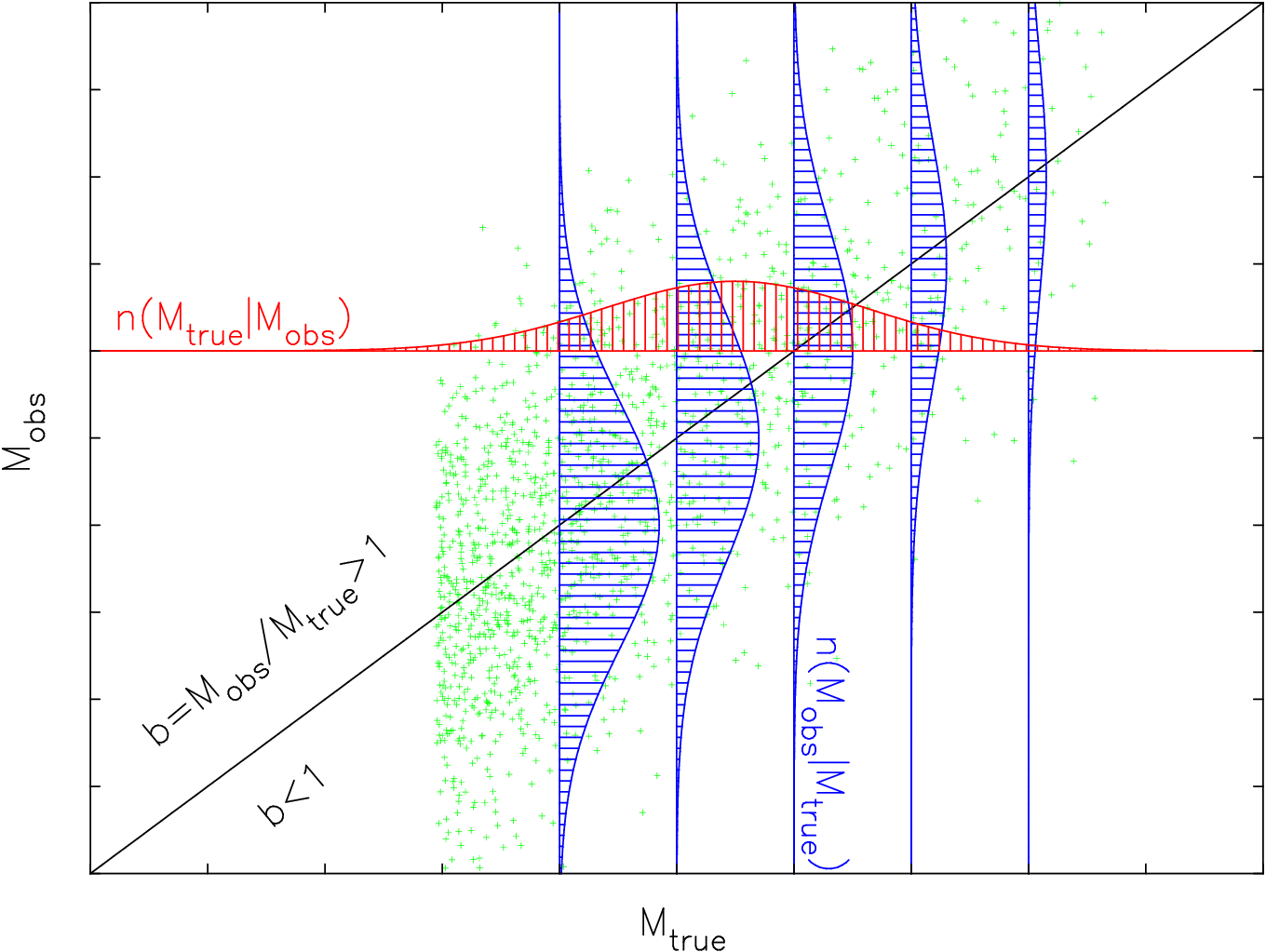}
\end{center}
\caption{A toy-model example of the excess up-scattering mass bias.
  Suppose one estimates weak lensing masses ($M_{\rm obs}$) of clusters of
  galaxies whose mass function decreases with
  increasing true mass ($M_{\rm true}$).
  The green dot symbols are an illustrative example of a distribution of
  clusters in $M_{\rm true}$--$M_{\rm obs}$ plane.
  Suppose errors in observed weak lensing masses are symmetric, the
  number counts of clusters with a fixed $M_{\rm true}$
  (or in a narrow $M_{\rm true}$--bin) as a function of 
  $M_{\rm obs}$ would look like $n(M_{\rm obs}|M_{\rm true})$ shown as
  blue vertical hatched distributions. 
  Due to the decreasing mass function, the number counts of clusters 
  with a fixed $M_{\rm obs}$ (or in a narrow $M_{\rm obs}$--bin) as a
  function of $M_{\rm true}$, $n(M_{\rm true}|M_{\rm obs})$, is not symmetric about
  $M_{\rm true}=M_{\rm obs}$, but would look like one shown as a red
  horizontal hatched distribution (note that this distribution can be
  skewed in actual cases). 
  It can be seen from it that the peak or mean of the distribution is
  located at a lower side of $M_{\rm true}=M_{\rm obs}$, which schematically
  explains the mechanism of the excess up-scattering mass bias.
  \label{fig:demo_exupscatbias}}
\end{figure}

Here we first give a briefly overview of the standard method of estimating cluster
mass, and then describe a mechanism of the bias arising in that process.
Commonly the standard $\chi^2$ analysis is adopted to estimate
individual cluster mass with the following procedure: 
A tangential shear profile, $\gamma_t(\theta)$, (or converting it into the
surface mass density, $\Delta\Sigma(R)$, see equation~(\ref{eq:gammat}))
is measured, and its theoretical  
model with a few model parameters (commonly, the cluster mass, $M$, and the
concentration parameter, $c$) is ready.
Then, along with covariance matrix in which appropriate noise components are
included, the $\chi^2$ distribution is computed, and is used to
derive point estimators (commonly the best likelihood point is taken)
and confidence intervals of model parameters.
Therefore, in short, a set of model parameters that, in a statistical
sense, best reproduces a measured {\it noisy} shear profile is taken as
the best-fitting model. 
It is indeed random noises in the measured shear profile combined
with a non-flat cluster mass function that induces the bias by the
mechanism described below (see also Figure~\ref{fig:demo_exupscatbias}
for a toy-model example):
A measured shear profile can be up-/down-scattered by
positive/negative noise, which may lead to over-/under-estimation of the
derived best-fitting weak lensing mass (which we denote as $M_{\rm obs}$).
Consider a cluster with an observed weak lensing mass of
$M_{\rm obs}^*$, its true cluster mass, $M_{\rm true}^*$, can be either
$M_{\rm true}^*<M_{\rm obs}^*$ or $M_{\rm true}^* > M_{\rm obs}^*$
depending on a noise realization. 
Since the cluster mass function is a decreasing function with
increasing mass, it is more likely that $M_{\rm obs}^*$ is the
result of up-scattering of smaller $M_{\rm true}^*$ than that of
down-scattering of larger $M_{\rm true}^*$.
Defining the mass bias parameter as $b=M_{\rm obs}/M_{\rm true}$, and
considering a
sample of clusters with observed weak lensing mass of $M_{\rm obs}^*$, the
above argument leads to an expectation that the averaged $b$ over the
sample is $\langle b \rangle>1$.
This is also the case for a sample of clusters with observed weak lensing
mass greater than $M_{\rm obs}^*$.
We call this bias ``{\it excess up-scattering mass bias}''.

The excess up-scattering mass bias is conceptually closely related to
the Eddington bias, which is known to have mainly two effects on weak
lensing cluster studies; the one is the mass bias described in the last
paragraph, and the other is the excess number counts of weak lensing
shear-selected clusters which are found as high peaks in weak lensing mass
maps \citep{2004MNRAS.350..893H,2012MNRAS.425.2287H,2018PASJ...70S..27M}.
The latter is caused by an excess up-scattering effect on weak
lensing peak signals in weak lensing mass maps reconstructed from noisy
shear data;
since weak lensing peak counts is a decreasing function
with increasing peak-height,
for a given observed weak lensing peak with its
height $\nu^*$, 
it is more likely that the $\nu^*$-peak is the result of up-scattering
of intrinsically (meaning an imaginary pure weak lensing signal without
noise) lower $\nu_{\rm int}^*(<\nu^*)$-peak than that of
down-scattering of intrinsically higher $\nu_{\rm int}^*(>\nu^*)$-peak.
\citet{2020ApJ...891..139C} examined Eddington bias on the weak lensing mass
estimate of weak lensing shear-selected clusters paying special
attention to the fact that weak lensing cluster finding and following
mass estimate use the same weak lensing shear data and thus both the
processes are affected by the up-scattering effect.
They quantified the mass biases for different cluster redshifts using a
large sample of realistic mock weak lensing clusters for a deep weak lensing
survey with a source galaxy number density of 30 arcmin${}^{-2}$.
\citet{2022A&A...661A..14R} studied X-ray properties of the
shear-selected clusters of \citet{2021PASJ...73..817O}, in which the
mass bias values derived in \citet{2020ApJ...891..139C} were 
used to correct the mass bias on the weak lensing cluster masses.

Recently, a weak lensing cluster search becomes a practical tool to construct a
large sample of clusters of galaxies thanks to dedicated wide
field surveys, for example, currently the largest sample is 187
  shear-selected clusters from the Hyper Suprime-Cam survey S19A data by
  \citet{2021PASJ...73..817O}.
In the near future, the size of shear-selected sample will
become much larger as many more wide-area weak lensing-oriented surveys will
come; in particular, the Legacy Survey of Space and
Time \citep[LSST,][]{2019ApJ...873..111I} and Euclid survey
\citep{2012SPIE.8442E..0TL,2018cosp...42E2761R} will cover a large
portion of the sky with a sufficient depth.
It is thus worth developing methods for correcting the mass bias 
on weak lensing mass estimate for shear-selected cluster samples without
relying on mock simulations of realistic weak lensing clusters.
This is exactly the purpose of this paper.

In this paper, we present an empirical method for correcting the
excess up-scattering mass bias on weak lensing mass estimate for
shear-selected cluster samples.
In doing this, we adopt the framework of the standard
Bayesian statistics along with the standard weak lensing cluster
mass estimate scheme, and we use recent models of properties of dark
matter halos under the cold dark matter structure formation scenario for
a prior knowledge of weak lensing clusters.
We test the method using mock simulations of weak lensing clusters.

The structure of this paper is as follows:
In Section~\ref{sec:basics_models}, we summarize basic theory of cluster
weak lensing and halo models of clusters of galaxies which are
used in weak lensing cluster mass estimate and to develop an
empirical method for correcting the excess up-scattering mass bias. 
In Section~\ref{sec:mock}, we describe a mock simulation of weak lensing
clusters which is used to examine excess up-scattering mass bias, and to
test its correction method.
Methods of weak lensing analyses of mock clusters and a method for
correcting the excess up-scattering mass bias are described in
Section~\ref{sec:analyses}.
Results of those analyses are presented in Section~\ref{sec:results}.
In Section~\ref{sec:HWL16a}, we apply our correction method to
the weak lensing shear-selected cluster sample of \citet{2020PASJ...72...78H}, and
present the bias corrected masses.
Finally, we summarize and discuss our results in
Section~\ref{sec:summary}. 

%%%%%% Sec-2: Basics and models of cluster weak lensing
\section{Basics and models of cluster weak lensing}
\label{sec:basics_models}

Here we summarize the basic theory of cluster weak lensing and halo models
of clusters of galaxies which are used in weak lensing cluster mass estimate
and to develop an empirical method for correcting the excess up-scattering
mass bias on the weak lensing mass estimate in section~\ref{sec:correction:method}.

%%%%%% Sec-2.1: Basics and models of cluster weak lensing
\subsection{Basics of cluster weak lensing}
\label{sec:basics}

The central quantity of cluster weak lensing is the tangential
component of the lensing shear at a source galaxy position
$\bm{\theta_s}$ relative to a cluster center $\bm{\theta_{\rm cl}}$,
$\gamma_t(\bm{\theta_s}:\bm{\theta_{\rm cl}})$.
This is, however, a noisy quantity for individual source galaxy; for
example, the amplitude of lensing 
shear from a cluster of our interest ($M \gtsim 10^{14}M_{\odot}/h$ at
$0.1\ltsim z \ltsim 0.8$) at
the cluster scale radius is $\sim 0.1$, but the root-mean-square of
intrinsic galaxy shape noise is $\sim 0.3$ per component.
Therefore, an optimal statistical estimator derived by an averaging
operation which efficiently extracts the cluster lensing signal while
reducing noise is employed according to a science case. 

For the cluster mass estimate, the most commonly used estimator is
the azimuthally averaged tangential
shear profile which relates to the excess surface mass density,
$\Delta\Sigma$, as \citep{1995ApJ...449..460K}, 
\begin{equation}
\label{eq:gammat}
\gamma_t(R) = {{\bar{\Sigma}(<R) - \Sigma(R)} \over
  {\Sigma_{\rm cr}(z_{\rm cl},z_{\rm s})}}
\equiv {{\Delta \Sigma(R)} \over
  {\Sigma_{\rm cr}(z_{\rm cl},z_{\rm s})}},
\end{equation}
where $\Sigma(R)$  and $\bar{\Sigma}(<R)$ are the azimuthally averaged
surface mass density at $R$ and interior to $R$, respectively, and 
the reciprocal critical surface mass density is calculated as
\begin{eqnarray}
\label{eq:sigmacr}
\Sigma_{\rm cr}^{-1}(z_{\rm cl},z_{\rm s}) =
\left\{
\begin{array}{ll}
0 & \mbox{if $z_{\rm s}\leq z_{\rm cl}$,}\\
{{4\pi G}\over {c^2}}
{{D_l D_{ls}} \over {D_s}} & \mbox{otherwise,}
\end{array}
\right.
\end{eqnarray}
where $D_l$, $D_s$ and $D_{ls}$ are the angular diameter distances
between an observer and a lens, between an observer and a lens, and
between a lens and a source, respectively.

For weak lensing cluster search, high peaks in a weak lensing mass map,
which is the convergence field ($\kappa$) convolved with a kernel
function ($U$), are searched for as strong candidates of massive clusters.
A weak lensing mass map is evaluated from the tangential shear data
\citep{1996MNRAS.283..837S}, 
\begin{eqnarray}
\label{eq:massmap}
{\cal{K}}(\bm{\theta}) &=& \int d^2 \bm{\phi}~
\kappa(\bm{\phi}-\bm{\theta}) U(|\bm{\phi}|)\nonumber\\
&=& \int d^2 \bm{\phi}~
\gamma_t(\bm{\phi}:\bm{\theta}) Q(|\bm{\phi}|),
\end{eqnarray}
where the filter function $Q$ is related to $U$ by,
\begin{equation}
\label{eq:U2Q}
Q(\theta) = {2\over \theta^2} \int_0^{\theta} d\theta' \theta'
U(\theta') -U(\theta).
\end{equation}
The truncated Gaussian kernel for $U$ is widely adopted
\citep{2002ApJ...580L..97M,2015PASJ...67...34H}, its $Q$ filter is 
\begin{equation}
\label{eq:Qfilter}
Q(\theta) = {1\over {\pi \theta^2}}
\left[
  1-\left(1+{{\theta^2} \over {\theta_G^2}}\right)
\exp\left( -{{\theta^2} \over {\theta_G^2}}\right)
\right],
\end{equation}
for $\theta<\theta_o$ and $Q=0$ elsewhere.
For filter parameters, we take $\theta_G=1.5$ arcmin and $\theta_o=15$
arcmin that were adopted in the weak lensing cluster finding by
\citet{2020PASJ...72...78H}.
The variance of noises on mass maps coming from the intrinsic galaxy
shapes (which we call the shape noise) is evaluated by
\citep{1996MNRAS.283..837S} 
\begin{equation}
\label{eq:sigmashape}
\sigma_{\rm shape}^2 = {\sigma_e^2 \over {2 n_g}}
\int d\theta \theta Q^2(\theta),
\end{equation}
where $\sigma_e$ is the root-mean-square value of intrinsic ellipticity
of galaxies and $n_g$ is the number density of source galaxies. 

%%% Sec-2.2: Halo models of clusters of galaxies
\subsection{Halo models of clusters of galaxies}
\label{sec:HaloModels}
For the model of the matter density profile of clusters of galaxies, we
adopt the truncated NFW model
\citep{1997ApJ...490..493N,2009JCAP...01..015B}, which is known to 
reproduce the averaged tangential shear profile of dark matter
halos measured from simulations of gravitational lensing ray-tracing through
cosmological dark matter distributions \citep{2011MNRAS.414.1851O} well.
%Its analytic expression of the tangential shear profile is given in
%\citet{2009JCAP...01..015B}. 
The NFW model is characterized by two parameters, the characteristic
density ($\rho_S$) and the scale radius ($r_s$) or equivalently the halo
mass and the concentration parameter.
To define those quantities, we take the conventional
over-density parameter ($\Delta$) relative to the critical density,  
$M_\Delta \equiv 4\pi r_{\Delta}^3 \Delta \rho_{cr}(z) /3 = M_{\rm NFW}(<r_\Delta)$.
The corresponding concentration parameter is given by
$c_\Delta = r_\Delta/r_s$.
For choices of the over-density, we take
$\Delta=200$ and 500, and denote the corresponding mass as $M_{200c}$
and $M_{500c}$, respectively.

For the halo mass function, we adopt a model by
\cite{2010ApJ...724..878T}.
For the model of a probability distribution of the concentration
parameter, we take the log-normal model,
\begin{equation}
\label{eq:pofc}
P(c|M)={1\over {\sqrt{2 \pi} \sigma_{\ln c}}}
\exp \left( -{{\left[\ln c - \ln \bar{c}(M)\right]^2} \over {2 \sigma_{\ln c}^2}}
\right){1\over c},
\end{equation}
where $\bar{c}(M)$ the mean mass-concentration relation for which we adopt a model by
\citet{2019ApJ...871..168D}, and we adopt $\sigma_{\ln c}=0.25$
following \citet{2020ApJ...891..139C}. 

%%% Sec-2.3: Projection effect by large-scale strucures
\subsection{Projection effect by large-scale structures}
\label{sec:lss}

In addition to weak lensing signals from a target cluster of galaxies, 
other lensing signals come from large-scale structures along the same
line-of-sight, which is called the projection effect. 
In order to include this effect into mock weak lensing clusters, we use
full-sky gravitational lensing simulation data from
\citet{2017ApJ...850...24T}. 
They performed gravitational lensing ray-tracing simulations using the
multiple-lens spherical plane algorithm through a nested system of
cubic $N$-body simulation boxes of different sizes
\citep{2015MNRAS.453.3043S}, which enables a full-sky gravitational
lensing simulation with a high angular resolution of 
$\theta_{\rm eff} <1$ arcmin. 
Lensing shear and convergence data are computed at grid points of the HEALPix 
pixelization with $N_{\rm side} =8192$ (corresponding to an effective pixel
scale of 0.43 arcmin, \citet{2005ApJ...622..759G}) on 38 source planes
located at equal intervals of the comoving 150 $h^{-1}$Mpc from $z=0.05$ to
$z=5.34$.
108 realizations of a full-sky lensing data set along with dark matter
halo catalog in the corresponding $N$-body boxes were generated.

In the full-sky simulation data, there are regions affected by lensing of
massive halos. 
In order to avoid a chance overlap between them and our mock cluster,
we exclude regions within 10 arcmin from halos with
$M_{200c} \ge 3\times 10^{14}M_{\odot}h^{-1}$, and within 5 arcmin from halos
with $8\times 10^{13}\le M_{200c}<3\times 10^{14}M_{\odot}h^{-1}$.
About 50 percent of the total area is excluded by these conditions.

%
%%%%%% Sec-3: Mock simulations of cluster weak lensing %%%%%%%%%%%%%%%%%%%%%%%%%%%%%%%%%%%%%%%%%%%%%%%%%%
%
\section{Mock simulations of cluster weak lensing}
\label{sec:mock}

Here we describe mock simulations of weak lensing clusters which are used
to examine the excess up-scattering mass bias in weak lensing
mass estimates and to test our method for correcting it.
In designing the mock simulation, we
adopt observational parameters of \citet{2020PASJ...72...78H} as we apply our bias
correction method to their weak lensing shear-selected cluster sample in
Section~\ref{sec:HWL16a}. 

%%% Sec-3.1: Setting of the mock simulation
\subsection{Setting of the mock simulation}
\label{sec:mock:setting}

Here we describe the setting of the mock simulation of a weak lensing cluster
observation. 

%%% Sec-3.1.1: Cosmological model
\subsubsection{Cosmological model}
\label{sec:mock:cosmology}
We adopt the WMAP cosmology \citep{2013ApJS..208...19H} as it was
adopted in the full-sky gravitational lensing simulations by
\citet{2017ApJ...850...24T} which we use to extract lensing shear signals
from large-scale structures in the same line-of-sights of mock clusters
(see Sections~\ref{sec:lss} and \ref{sec:mock:recipe}). 

%%% Sec-3.1.2: Redshift distribution of source galaxies
\subsubsection{Redshift distribution of source galaxies}
\label{sec:mock:ns}

%
%%%%% Fig wsumpdf_0z3_pth095 --- fig-2
%
\begin{figure}
\begin{center}
  \includegraphics[width=82mm]{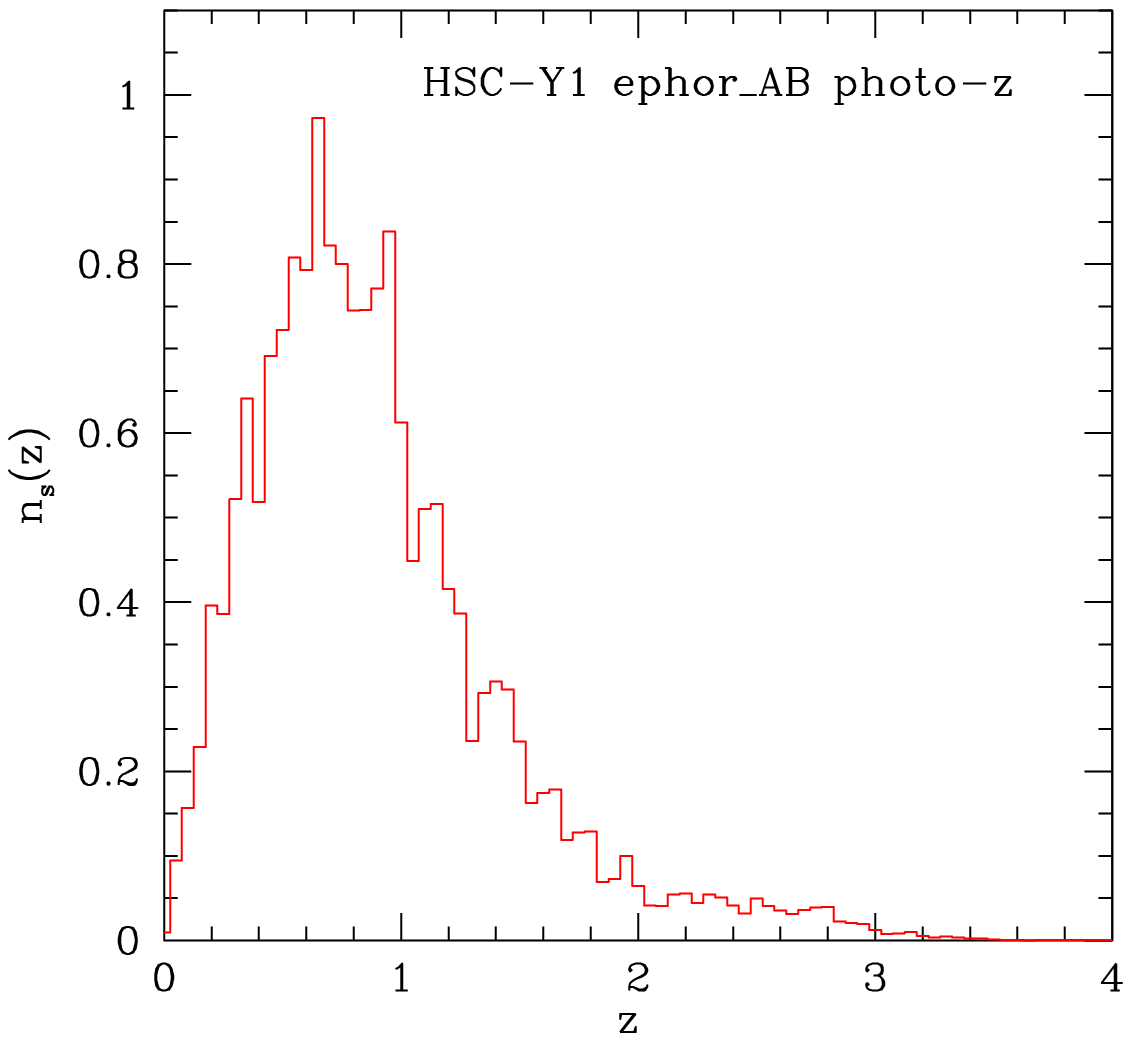}
\end{center}
\caption{An estimate of redshift distribution of the
  source galaxy sample used in a weak lensing cluster search by
  \citet{2020PASJ...72...78H}, which is adopted in simulations of mock
  weak lensing clusters in this study.
  The distribution is computed by summing up the redshift probability
  distributions, which are derived in photo-$z$ estimation for each
  galaxy, over selected source galaxies.
  The normalization is taken so that $\int dz n_s(z)=1$
  \label{fig:ns}}
\end{figure}

We adopt the redshift distribution of source galaxies estimated by 
\citet{2020PASJ...72...78H} shown in Figure~\ref{fig:ns}.
In short, they performed a weak lensing cluster search using the Hyper
SuprimeCam survey first-year (HSC-Y1) weak
lensing shape catalog \citep{2018PASJ...70S..25M}, to which photometric
redshift (photo-$z$) information derived using HSC five-band photometry
\citep{2018PASJ...70S...9T} is linked.
They adopted {\tt Ephor\_AB} photo-$z$ data among six photo-$z$ methods
applied to the HSC-Y1 galaxy catalog \citep{2018PASJ...70S...9T}, and
selected source galaxies using the photo-$z$ information with the $P$-cut
method \citep{2014MNRAS.444..147O}.
The redshift distribution of selected galaxies shown in
Figure~\ref{fig:ns} is estimated by summing
the full probability distribution functions of redshift, $P(z)$, derived
in photo-$z$ estimations taking into account the lensing weight $w$; 
$n_s(z)= \sum_i w_i P_i(z)/\sum_i w_i$, 
where the summation is taken over all the selected galaxies.
The effective number density of selected galaxies was
$n_g=19.3$ arcmin$^{-2}$.

%%% Sec-3.2: The recipe for mock weak lensing clusters
\subsection{The recipe for mock weak lensing clusters}
\label{sec:mock:recipe}

We generate samples of mock weak lensing clusters for four cluster
redshifts, $z_{\rm cl} = 0.15$, 0.25, 0.35, and 0.45.
Each mock cluster and associated source galaxies are generated by the
following procedure: 

(1) {\it Random choices of a true cluster mass and concentration:}
We consider the range of the cluster mass
$1\times 10^{13}<M_{200c}<2\times 10^{15}M_\odot h^{-1}$,
and randomly choose a true
mass (which we denote $M_{\rm true}$) according to the halo
mass function by \cite{2010ApJ...724..878T}. 
Then, we randomly choose a concentration parameter according to the
log-normal probability distribution, equation~(\ref{eq:pofc}).
  
(2) {\it Adding a perturbation to the cluster mass to take account of
  complexities of cluster mass distribution:}
Mass distributions of clusters do not exactly follow the NFW profile,
but deviate from it due to, for example, the triaxiality and existence of
substructures
\citep{2005ApJ...632..841O,2007MNRAS.380..149C,2010A&A...514A..93M,2011ApJ...740...25B,2011MNRAS.414.1851O,2012MNRAS.421.1073B}.
In order to include effects of these deviations into mock weak lensing
clusters, we follow a prescription by \citet{2020ApJ...891..139C}; that
is, we adopt a statistical relation between a realistic halo with a
true mass $M_{\rm true}$ and an NFW halo with a mass $M_{\rm NFW}$ whose
tangential shear profile best describes that of the realistic halo,
$P(M_{\rm NFW}|M_{\rm true})$.
Based on the findings by
\citet{2011ApJ...740...25B}, and \citet{2012MNRAS.421.1073B},
\citet{2020ApJ...891..139C}  modeled $P(M_{\rm  NFW}|M_{\rm true})$ by a
log-normal distribution, 
\begin{equation}
\label{eq:intrinsic_scatter}
M_{\rm  NFW}=M_{\rm true} \exp(\Delta_{\rm NFW}),
\end{equation}
where $\Delta_{\rm NFW}$ is the normal distribution with the standard
deviation of $\sigma_{\rm NFW}$.
\citet{2020ApJ...891..139C} took $\sigma_{\rm NFW}=0.18$ which we 
adopt in this work, which
approximately corresponds to a 20 percent scatter in mass.
Thus, for each mock cluster with $M_{\rm true}$, we draw an
$M_{\rm NFW}$ according to equation~(\ref{eq:intrinsic_scatter}), and
use it to model the lensing shear profile later in step (4). 

(3) {\it Placing source galaxies in the sky and redshift-space:}
Source galaxies are randomly placed around a cluster
with the mean number density of $n_g=20$ arcmin$^{-2}$, and a redshift
of each galaxy is assigned randomly according to the redshift probability
distribution of \citet{2020PASJ...72...78H} shown in Figure~\ref{fig:ns}
(see section~\ref{sec:mock:ns} for details). 
The redshift range is limited to $0<z<5.34$ because of 
the availability of the full-sky
gravitational lensing simulation data (see Section~\ref{sec:lss}), 

(4) {\it Assigning shear signals and noise to each galaxy:}
Having set the model parameters of a cluster and positions and redshift
of the source galaxies, we assign weak lensing shear signals and
a noise to each galaxy.
We take into account the following three components:
Lensing shear signals from the cluster ($\gamma_{\rm cl}$) and from large-scale
structures in the same line-of-sight of each source galaxy ($\gamma_{\rm lss}$,
see section~\ref{sec:lss}), and a noise due to intrinsic galaxy
shape ($n_{\rm shape}$).
We describe them in turn as follows:
\renewcommand{\labelenumi}{(\roman{enumi})}
\begin{enumerate}
\item Cluster component: We adopt the truncated NFW model (see
  section~\ref{sec:HaloModels}) for which the analytical expressions of
  the lensing signals are given in the literature \citep[e.g.,
  ][]{2009JCAP...01..015B}. Therefore, for each source galaxy with given
  redshift and perpendicular distance from a cluster center, the
  cluster shear component, $\gamma_{\rm cl}$, is calculated using the
  analytical expressions. 
\item Large-scale structure component: We use full-sky gravitational
  lensing simulation data by \citet{2017ApJ...850...24T} (see
  section~\ref{sec:lss} for its brief description) to assign
  $\gamma_{\rm lss}$ to each galaxy: First, for each cluster, we
  randomly choose its position in a simulation sky, 
  and then the sky position of every source galaxy is automatically
  determined and the simulation pixel containing it is as well.
  Then for each source galaxy, we extract simulation shear data from the
  corresponding pixel on the source plane which is closest to the galaxy
  redshift and assign it to that galaxy.
\item Noise component: For each galaxy, we draw an intrinsic shape noise
  $n_{\rm shape}$ according to the normal distribution with the standard
  deviation of $\sigma_{\gamma,{\rm int}}=0.4$.   
\end{enumerate}
For each cluster redshift, we generate 10,000 clusters per one full-sky
simulation data set. We used all the 108 available data sets, and thus
we have 1,080,000 clusters for each cluster redshift.

%%% Sec-4: Weak lensing anayses of mock clusters
\section{Methods of weak lensing analyses and a mass bias correction}
\label{sec:analyses}

Having prepared the mock weak lensing shear catalog for each
cluster, we first perform two weak lensing analyses, a measurement of
weak lensing peak height in a mass map, and a standard weak lensing
cluster mass estimate using a tangential shear profile.
We then test our empirical method for mitigating the 
excess up-scattering mass bias on the weak lensing mass estimate for
shear-selected cluster samples.
In this section, we describe methods for these analyses.

%%% Sec-4.1: The peak height in a weak lensing mass map
\subsection{The peak height in a weak lensing mass map}
\label{sec:mock:peakheight}

The peak height of a cluster in a weak lensing mass map is computed by
a discrete form of equation~(\ref{eq:massmap}),
\begin{equation}
\label{eq:peakheight}
{\cal{K}} = {1\over {n_g}} \sum_i \gamma_{t,i} Q(|\bm{\phi}_i|),
\end{equation}
where the summation is taken over all galaxies within $\theta_o$ from
a cluster center, and the filter function $Q$ is given
by equation~(\ref{eq:Qfilter}). We evaluate the shape noise
$\sigma_{\rm shape}^2$ using equation~($\ref{eq:sigmashape}$) with
$n_g=20$arcmin$^{-2}$ and $\sigma_e=0.4$.
We define the signal-to-noise ratio of the peak height as
\begin{equation}
\label{eq:SNpeak}
SN_{\rm peak}={\cal{K}}/\sigma_{\rm sharp}.
\end{equation}
In addition to this noisy estimator, we also compute $SN_{\rm NFW}$ 
which is a peak signal solely from the NFW cluster shear (by replacing
$\gamma_t=\gamma_{t,{\rm cl}}+\gamma_{t,{\rm lss}}+n_{t,{\rm shape}}$ in
equation~(\ref{eq:peakheight}) with $\gamma_{t,{\rm cl}}$) normalized by
the same $\sigma_{\rm shape}$.

%%% Sec-4.2: Standard weak lensing cluster mass estimate
\subsection{Standard weak lensing cluster mass estimate}
\label{sec:mock:WLmass}

%
%%%%% Fig z25_run0574paper.ps --- fig-3
%
\begin{figure}
  \begin{center}
  \includegraphics[height=82mm,angle=270]{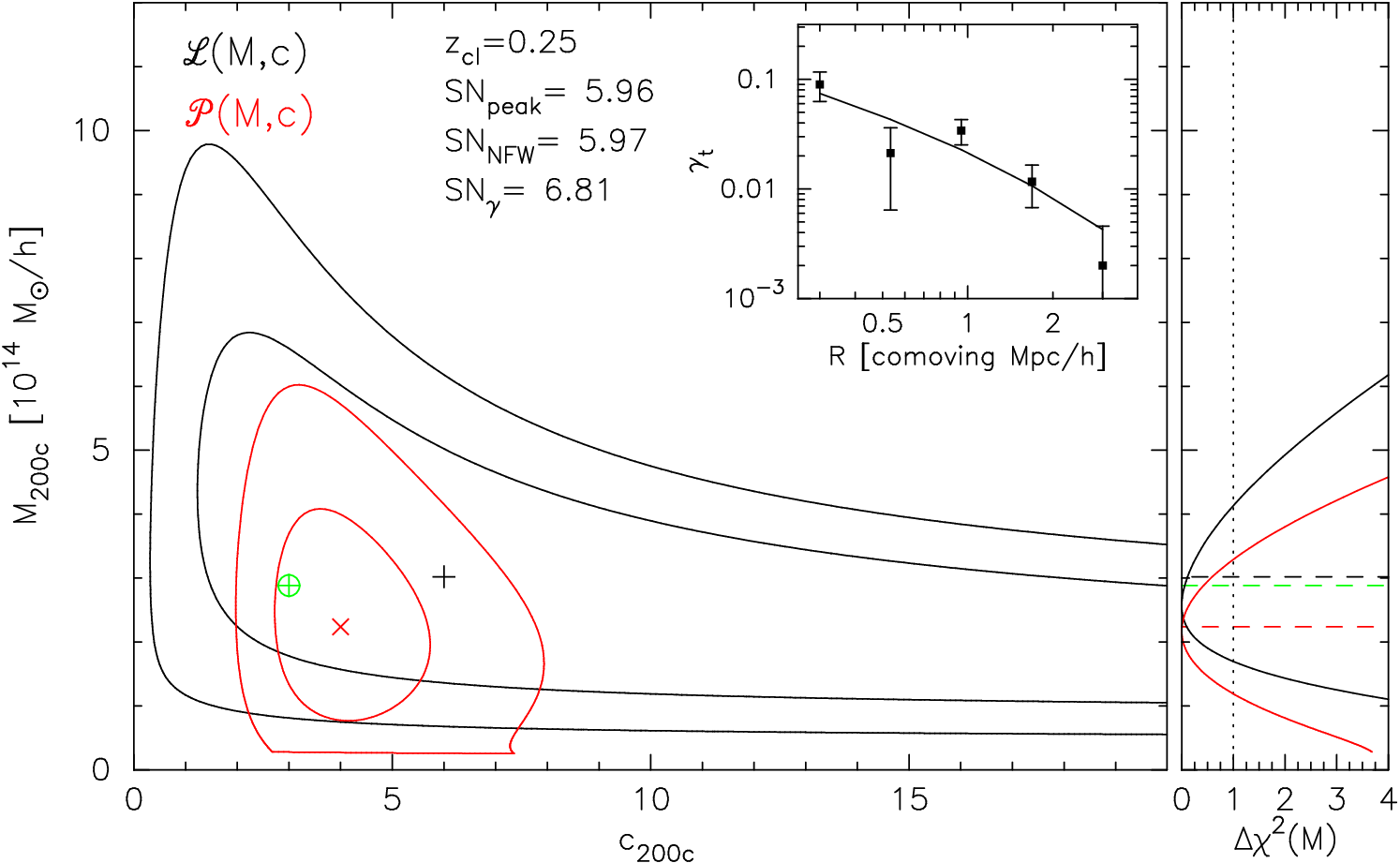}
\end{center}
\caption{Example of mock tangential shear profile and results of
  $\chi^2$-fit of model parameters. Results of a mock cluster of
  $z_{\rm cl} = 0.25$  are shown.
  In the inserted panel, points show bin-averaged tangential shear
  profile ($\gamma_t(R_i)$) with error bars showing
  diagonal components of covariance matrix ($\sqrt{\rm Cov}_{ii}$).
  The solid line shows the best fit model.
  The left panel show contour maps (68\% and 95\% confidence levels)
  of the likelihood (black) and posterior (red) functions with best-fitting
  positions being marked by the plus and cross, respectively.
  The green plus in a circle shows the input model parameters.
  The right-hand panel shows the marginalized one-dimensional probability
  functions of the cluster mass in the form of
  $\Delta \chi^2(M)=-2\log{\cal{L}}(M)/{\cal{L}}_{\rm best}$.
  Dashed lines show the true mass and best-fitting masses shown in the left-hand
  panel.  
  \label{fig:mock_example}}
\end{figure}

In short, we derive a cluster mass by first computing the radial
bin-averaged tangential shear profile for each cluster, and then fitting
it to the NFW model profile based on the standard $\chi^2$ likelihood
analysis. Below we describe those steps in more details
\citep[see][and references therein for further
  details]{2019ApJ...875...63M,2020ApJ...890..148U}. 

We compute a bin-averaged tangential shear profile in five radial 
bins of equal logarithmic spacing of $\Delta \log R=0.25$ with bin
centers of
$R_c(i)=0.3\times 10^{i \Delta \log R}$ [$h^{-1}$Mpc], where $i$ runs from
0 to 4.
In order to avoid the dilution effect by foreground and/or cluster
member galaxies
\citep{2005ApJ...619L.143B,2007ApJ...668..643L,2007MNRAS.379..317H,2018PASJ...70...30M,2008ApJ...684..177U,2010PASJ...62..811O}, 
we use only galaxies that are well behind a target cluster, to be specific,
$z>z_{\rm cl}+0.05$.

We employ standard likelihood analysis for deriving constraints on 
model parameters.
The log-likelihood is given by
\begin{equation}
\label{eq:loglike}
-2 \ln {\cal{L}}(\bm{p}) =
\sum_{i,j} [d_i -m_i(\bm{p})] {\rm Cov}_{ij}^{-1} [d_j -m_j(\bm{p})],
\end{equation}
where the data vector $d_i=\gamma_t(R_i)$, and $m_i(\bm{p})$ is the
model prediction with the model parameters
$\bm{p}=(M_{200c},c_{200c})$.
%%% rev202209
We adopt the truncated NFW model (see Section \ref{sec:HaloModels})
for the lens model, whereas deviations from it due to, e.g.,
triaxiality and existence of substructures, are taken into account in
the covariance (see below).
In computing the source redshift weighted tangential shear profile of
the truncated NFW model, we use the redshift distribution of source
galaxies for each cluster.  
The covariance matrix (Cov) is composed of the three components,
\citep[see][and references therein for detailed descriptions]{2020ApJ...890..148U}: 
The statistical uncertainties due to the galaxy intrinsic shape
noise (Cov$^{\rm shape}$); the cosmic shear covariance due to the
projection effect of uncorrelated large-scale structures
\citep[Cov$^{\rm lss}$, see for details][]{2003MNRAS.339.1155H}; and the intrinsic
variance of the cluster lensing signal due to, e.g., triaxiality of
clusters and presence of substructures
\citep[Cov$^{\rm int}$, see for details][]{2015MNRAS.449.4264G,2019ApJ...875...63M}.

We compute the log-likelihood function over the two-parameter space in
the range of
$5\times 10^{12}<M_{200c}< 3\times 10^{15} M_{\odot}h^{-1}$, and $0.1<c_{200c}<30$.
We adopt the maximum point of the likelihood function as the best-fitting model,
and denote its mass as $M_L$.
In order to derive confidence intervals of $M_{200c}$, we marginalize over
$c_{200c}$ in the range $1<c_{200c}<30$ with $1/c_{200c}$ weight (or
a log-uniform prior for $c_{200c}$), because it is appropriate to assume a
log-uniform prior, instead of a uniform prior, for a positive-definite
quantity \citep{2020ApJ...890..148U}, and the lower-bound of
$c_{200c}>1$ is taken as a physically reasonable choice. 

In order to define the signal-to-noise ratio of hte measured tangential
shear profile, we employ the standard quadratic estimator, 
\begin{equation}
\label{eq:sngamma}
SN_\gamma^2 = \sum_{i} \left[ \gamma_t(R_i) /  {\rm Cov}_{ii}^{\rm shape}\right]^2,
\end{equation}
where the summation is taken over all the bins with $\gamma_t(R_i)>0$.

For the pirpose of illustration, we present one example of weak lensing
analyses of a mock cluster in Figure~\ref{fig:mock_example}.
Shown is a case with 
$M_{\rm true} \simeq 3\times 10^{14}M_\odot h^{-1}$ (in $M_{200c}$) and
$z_{\rm cl}=0.25$.
As one can see in the plot, for most individual clusters, the
concentration parameter is not well constrained due to a limited $R$
range.
Also it is unlikely that estimated concentration parameters are affected by a
systematic bias like the excess up-scattering mass bias, because, unlike
the halo mass function, the probability distribution of the
concentration parameter is not a monotonic function. 
Therefore, in this paper we do not deal with constraints on the
concentration parameter.

%
%%%%%% Sec-4.3: An empirical method for mitigating the excess up-scattering mass bias
%
\subsection{An empirical method for mitigating the excess up-scattering mass bias}
\label{sec:correction:method}

In developing a method, we have set the following two requirements:
(1) It should not rely on simulations of mock weak lensing clusters, but 
should be based on analytical or empirical models of dark matter halos
of cluster scale.
(2) It can be directly incorporated into the standard weak lensing
cluster mass estimation scheme.

For the framework of our correction method, we employ the standard
Bayesian statistics written in conventional notation as follows:
\begin{equation}
\label{eq:posterior}
P(M,c|\gamma_t^{obs}) \propto L(\gamma_t^{obs}|M,c) Pr(M,c),
\end{equation}
where
$P$, $L$, and $Pr$ are the posterior probability, the likelihood
function, and the prior, respectively, with model parameters $(M,c)$ and 
an observed shear profile ($\gamma_t^{obs}$).

A natural choice of the likelihood function is ${\cal{L}}(M,c)$
defined in equation~(\ref{eq:loglike}).
For the prior of the cluster mass and concentration parameter, we employ
empirical models of their probability distributions from recent
numerical simulations of structure formation; the mass function of
dark matter halos, $n(M)$, by \citet{2010ApJ...724..878T}, and the
log-normal probability distribution of the concentration parameter,
equation~(\ref{eq:pofc}), with the mean mass--concentration relation
by \citet{2019ApJ...871..168D}, thus we have
\begin{eqnarray}
\label{eq:prior}
Pr(M,c) = \left\{
\begin{array}{ll}
n(M) P(c|M) & \mbox{for $SN_\gamma^{\rm NFW}(M,c)>1$,}\\
0 & \mbox{otherwise,}
\end{array}
\right.
\end{eqnarray}
where $SN_\gamma^{\rm NFW}(M,c)$ is the signal-to-noise ratio of shear
profile of NFW models defined in the same manner as in 
equation~(\ref{eq:sngamma}) but using a model prediction of the shear
profile.
The condition of $SN_\gamma^{\rm NFW}(M,c)>1$ is imposed to avoid 
the prior increasing monotonically toward lower mass, which results
in a non-physical upturn of the posterior at a very low-mass range.
The choice of threshold is arbitrary, but it needs to include all the
reasonably possible parameter space. 
We empirically take the threshold of $SN_\gamma^{\rm NFW}>1$ considering
our selection of $SN_{\rm peak}>5$.

%%% Sec 4-4: Note on different $SN$s
\subsection{Note on different $SN$s}
\label{sec:noteSNs}

In this study, we use different $SN$s, namely, 
$SN_{\rm peak}$ (defined by equation (\ref{eq:SNpeak})),
$SN_{\rm NFW}$ (introduced in section~\ref{sec:mock:peakheight}),
$SN_{\gamma}$ (defined by equation (\ref{eq:sngamma})),
and $SN_{\gamma}^{\rm NFW}$ (introduced in
section~\ref{sec:correction:method}).
Here we clarify the differences between them.

$SN_{\rm peak}$ and $SN_{\gamma}$ are actual observables, and are
strongly correlated with each other because those $SN$s come from the
same lensing shear signals of a cluster, although different angular
weights (the kernel function $Q$ for a weak lensing mass map, and an
angular scale of $\gamma_t$ measurement) cause scatter between them.
In addition, in actual measurement processes, there is another factor
that may cause a systematic difference between them.
It is different selections of source galaxies used in each measurement:
In weak lensing peak search, the redshifts of clusters are unknown in
advance, and thus a galaxy sample used to generate a mass map is
commonly selected by a simple magnitude cut on a single photometry band
\citep[for example][]{2002PASJ...54..833M,2015PASJ...67...34H}.
Such a galaxy sample inevitably contains foreground and cluster member
galaxies, which results in the dilution effect on the peak heights
and thus on $SN_{\rm peak}$ \citep{2020PASJ...72...78H}.
Once a peak is matched with a known cluster with redshift information, a
sample of background galaxies can be selected by multi-band photometric
information \citep[see][ and references therein]{2018PASJ...70...30M},
which is used to estimate a weak lensing cluster mass with the
$\gamma_t$ profile. 
Therefore $SN_{\gamma}$ is less affected by the dilution effect.

$SN_{\rm NFW}$ and $SN_{\gamma}^{\rm NFW}$ are defined in the same
manner as $SN_{\rm peak}$ and $SN_{\gamma}$, 
respectively, but their signal levels are computed assuming a pure NFW
model without taking any noise component into account.
Thus, those $SN$s are model predictions of corresponding peak and $\gamma_t$
measurements for a cluster with a set of model parameters.

%%% Sec-5: Results of mock weak lensing cluster analyses
\section{Results}
\label{sec:results}

In this section, we present results of weak lensing analyses of mock
clusters.
We first present results of standard weak lensing
analyses (described in sections~\ref{sec:mock:peakheight} and
\ref{sec:mock:WLmass}) focusing on the excess up-scattering mass
bias.
Then, we test the method for correcting the bias described in
section~\ref{sec:correction:method}.

\subsection{Results of standard weak lensing analyses of mock clusters}
\label{sec:mock:results}

%
%%%%% Fig mtrue_mlike_snpeak_snlens_z0.25 --- fig-4
%
\begin{figure}
\begin{center}
  \includegraphics[height=82mm,angle=270]{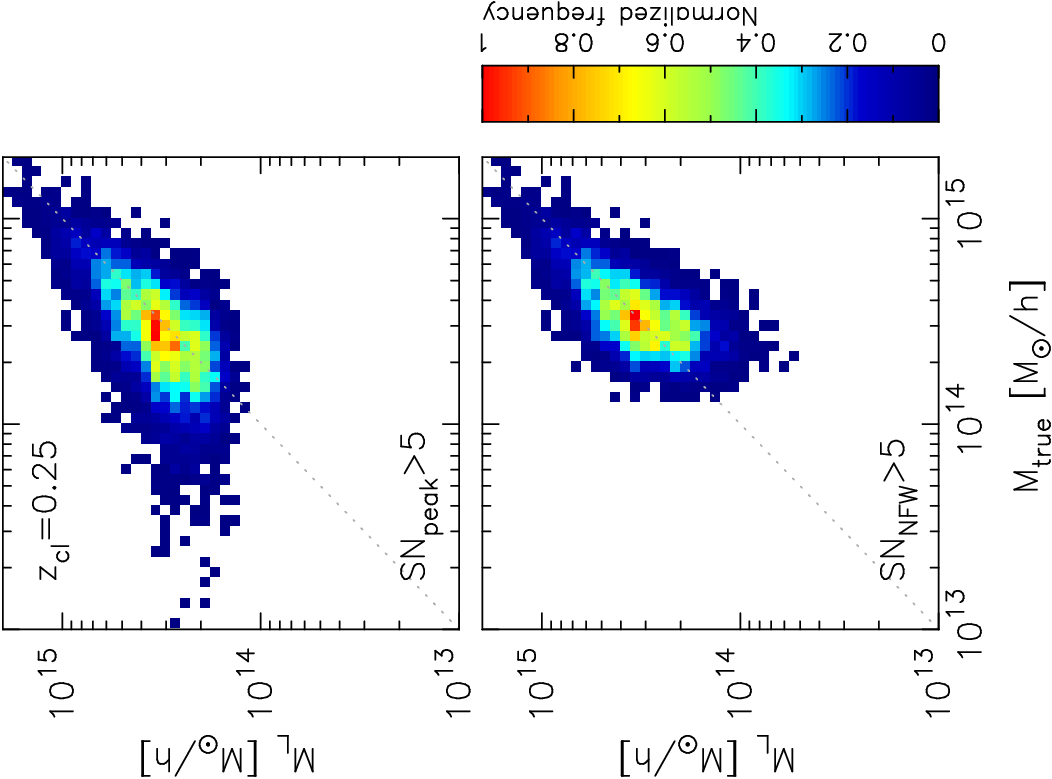}
\end{center}
\caption{Frequency distributions of weak lensing clusters in
  the $M_{\rm true}$--$M_L$ (in $M_{200c}$) plane.
  The top-panel is for clusters selected with the noisy weak lensing
  peak height of $SN_{\rm peak}>5$, whereas the bottom-panel is for
  clusters selected with the peak height only from the NFW
  component of $SN_{\rm NFW}>5$.
  The results from the mock cluster sample of $z_{\rm cl}=0.25$ are shown.
  \label{fig:mock_Mtrue-ML}}
\end{figure}

%
%%%%% Fig bias_mlike_snpeak_snlens --- fig-5
%
\begin{figure}
\begin{center}
  \includegraphics[width=82mm]{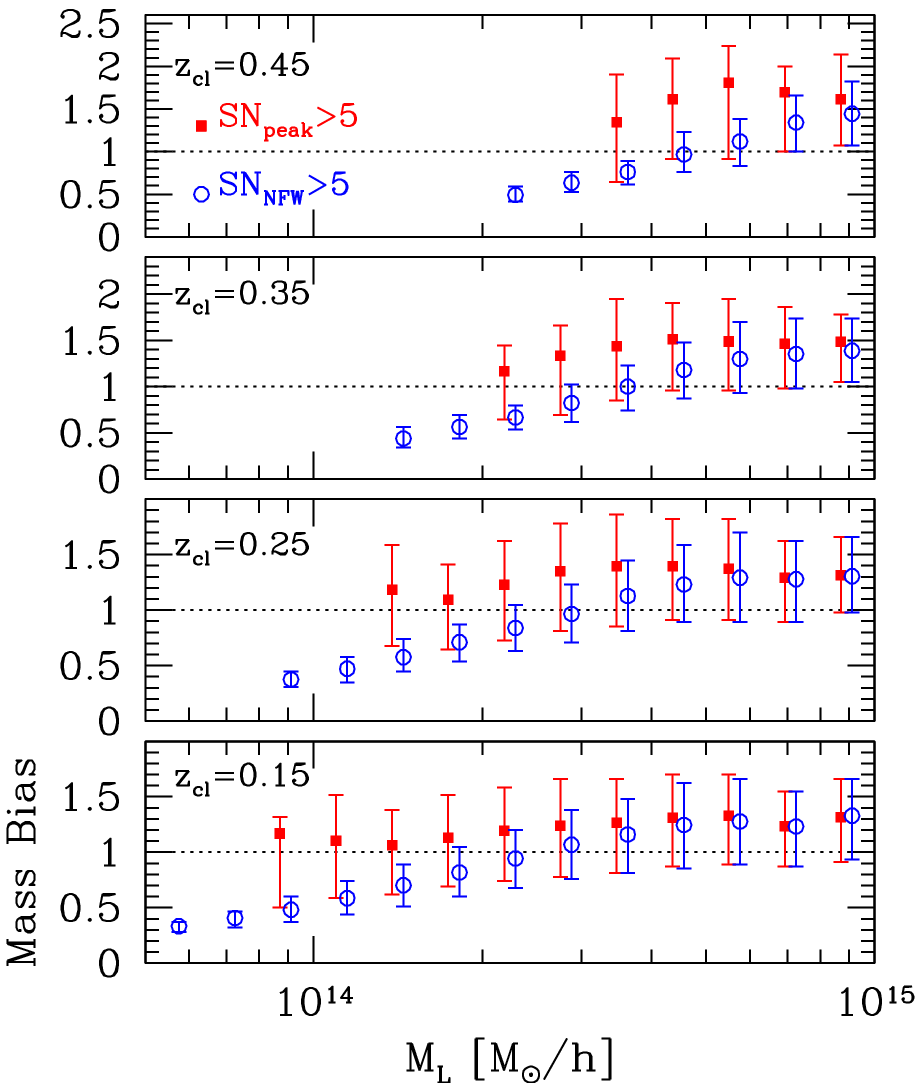}
\end{center}
\caption{Mass biases, defined by $b=M_L/M_{\rm true}$, as a function of
  $M_L$  (in $M_{200c}$) is
  shown. Symbols show the average values of mock cluster samples in $M_L$-bins, and error
  bars show the 16th and 84th percentiles of the samples in each bin.
  Filled squares are for clusters selected with the noisy weak lensing
  peak height of $SN_{\rm peak}>5$, whereas open circles are for
  clusters selected with the peak height only from the NFW
  component of $SN_{\rm NFW}>5$.
  Note that the results are slightly shifted in the horizontal direction
  for clarity. 
  \label{fig:mock:bias_mlike}}
\end{figure}

Figure~\ref{fig:mock_Mtrue-ML} shows frequency distributions of weak
lensing clusters in the $M_{\rm true}$--$M_L$ plane.
The top-panel is for clusters selected with the noisy weak lensing peak
height of $SN_{\rm peak}>5$ (namely, shear-selected clusters), whereas
the bottom-panel is for clusters
selected with the peak height from the NFW component only, 
$SN_{\rm NFW}>5$. 
In the bottom panel, there is a cut-off at
$M_{\rm true} \sim 1\times 10^{14}M_{\odot}h^{-1}$ which is due to a
tight correlation between $M_{\rm true}$  at a fixed $z_{\rm cl}$ and
$SN_{\rm NFW}$ combined with the $SN_{\rm NFW}$ threshold.
In the top panel, no such cut-off is seen, but the distribution spreads
toward lower $M_{\rm true}$, which is due to the Eddington bias studied
in detail by \citet{2020ApJ...891..139C};
in short, some fraction of clusters with $SN_{\rm NFW}$ values below the
selection threshold are up-scattered by noises and pass the selection
threshold of $SN_{\rm peak}>5$.
The tangential shear profiles of such clusters are also
most likely up-scattered, because the same
weak lensing shear data are used in both the peak height measurement and
the weak lensing mass estimate.
Therefore, as a result, for such up-scattered clusters, $M_L$ is
systematically larger than  $M_{\rm true}$.
This is quantitatively shown in Figure~\ref{fig:mock:bias_mlike}, where
the mass bias defined by
\begin{equation}
\label{eq:massbias}
b = {{M_L} \over {M_{\rm true}}},
\end{equation}
is plotted as a function of $M_L$, with symbols showing the average
values of mock cluster samples in $M_L$-bins, and error bars showing the
16th and 84th percentiles of the binned samples.
It is seen in the figure that for the sample with $SN_{\rm peak}>5$
selection, the average values of biases are greater than unity; to be
specific $\langle b\rangle \sim 1.1-1.8$ depending on $M_L$ and the
redshift. 

Another important point seen in Figure~\ref{fig:mock:bias_mlike} is that
even for the sample selected by $SN_{\rm NFW}>5$, the average values of biases
are greater than unity at high-$M_L$ ranges. 
Since there is no Eddington bias effect on this $SN_{\rm NFW}$-selected 
sample, this is purely the result of the excess up-scattering mass
bias.
This indicates that the excess up-scattering mass bias is not a
phenomenon unique to shear-selected samples but can also present in
cluster samples selected by other methods.
Notice that at lower $M_L$ ranges, the biases decline as
decreasing $M_L$. This is a selection effect caused by the low-$M_{\rm true}$
cut-off seen in the bottom-panel of Figure~\ref{fig:mock_Mtrue-ML}, and 
because of this, only down-scattered clusters can enter lower-$M_L$ bins,
resulting in $b<1$.

It is important to note that although the best-fitting masses are, on
average, biased high, the marginalized one-dimensional likelihood
distributions of the mass give a statistically correct confidence intervals.
In fact, it is found from results of mock weak lensing cluster analyses
of shear-selected sample ($SN_{\rm peak}>5$) that in 72(96) percent of
samples, $M_{\rm true}$ is within a conventional
1(2)-$\sigma$ confidence interval of the marginalized distribution (to be
more specific, $M_{\rm true}$ is within the range of
$\Delta\chi^2(M)<1(4)$).
This indeed suggests that the weak lensing mass estimate scheme itself
is correctly working, but the probability of up- and down-scattering is
not symmetric, causing the biased best-fitting masses, on average.

%%% rev202209
%
%%%%% Fig bias_mtrue_snpeak_snlens --- fig-6
%
\begin{figure}
\begin{center}
  \includegraphics[width=82mm]{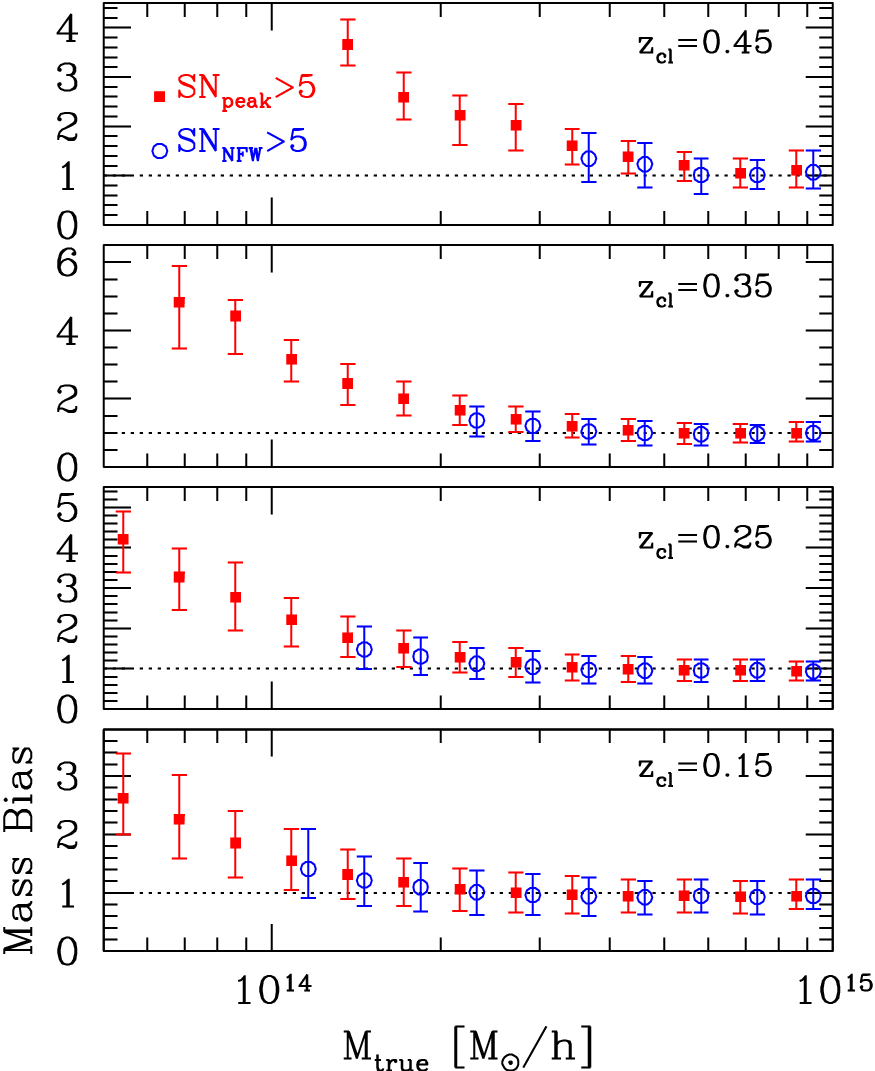}
\end{center}
\caption{Same as Figure~\ref{fig:mock:bias_mlike} but mass biases as a
  function of $M_{\rm true}$  (in $M_{200c}$) is shown.
  To be specific, the mass biases are computed for mock cluster samples in
  $M_{\rm true}$-bins.
  \label{fig:mock:bias_mtrue}}
\end{figure}

% rev202209 begin
Next, we look into the mass bias as a function of
$M_{\rm true}$, because, although it is not a direct observable, it may
give some insight into the excess up-scattering mass bias. 
Figure~\ref{fig:mock:bias_mtrue} shows the mass bias computed for mock
cluster samples in $M_{\rm true}$-bins.
At lower $M_{\rm true}$ ranges, the bias for the $SN_{\rm peak}>5$
selection increases as decreasing $M_{\rm true}$,
which is due to the Eddington bias effect seen in the top-panel of
Figure~\ref{fig:mock_Mtrue-ML}. The bias for $SN_{\rm NFW}>5$ selection
also shows a similar trend which is due to a selection effect as is seen in the
bottom-panel of Figure~\ref{fig:mock_Mtrue-ML}. 
At higher $M_{\rm true}$ ranges, averaged biases are very close to unity.
This is a natural consequence of nearly unbiased noise
properties of tangential shears, leading to nearly unbiased scattering
of $M_L$ for a sample of clusters within a narrow $M_{\rm true}$-bin.
It should be emphasized that the excess up-scattering mass bias
does not arise solely from up- and down-scattering due to unbiased
noises, but arises from scatterings over a wide mass range 
combined with a non-flat cluster mass function. 
Therefore, an averaged bias of a sample of clusters in a narrow
$M_{\rm true}$ range is not affected by the excess up-scattering
effect, 
because non-symmetric scatterings do not occur in that sample.
However, when one infers the true mass of a weak lensing cluster from a
wide range of potential true masses (or a mass range where 
a likelihood, ${\cal L}(M)$, is non-negligible), one needs to take into
account the non-symmetric scattering effect and thus needs to take into
account the shape of the mass function (or the prior information in
the Bayesian framework).
This is a basic concept of our method for correcting the excess
up-scattering mass bias. 

%
%%%%%% Sec-4.2: Test with mock weak lensing clusters
%
\subsection{Testing the bias correction method with mock weak lensing
  clusters}
\label{sec:correction:test}

We test our method for mitigating the excess up-scattering mass bias with
the mock weak lensing clusters.
%described in Section~\ref{sec:mock}.
For each mock cluster, we compute the posterior distribution defined in
equation~(\ref{eq:posterior}) with the likelihood function,
equation~(\ref{eq:loglike}), and the prior, equation~(\ref{eq:prior}).
We take the maximum point of the posterior distribution as the best-fitting
model, and denote its mass as $M_P$.

%
%%%%% Fig fract_mpost_mlike_snds --- fig-7
%
\begin{figure}
\begin{center}
  \includegraphics[width=82mm]{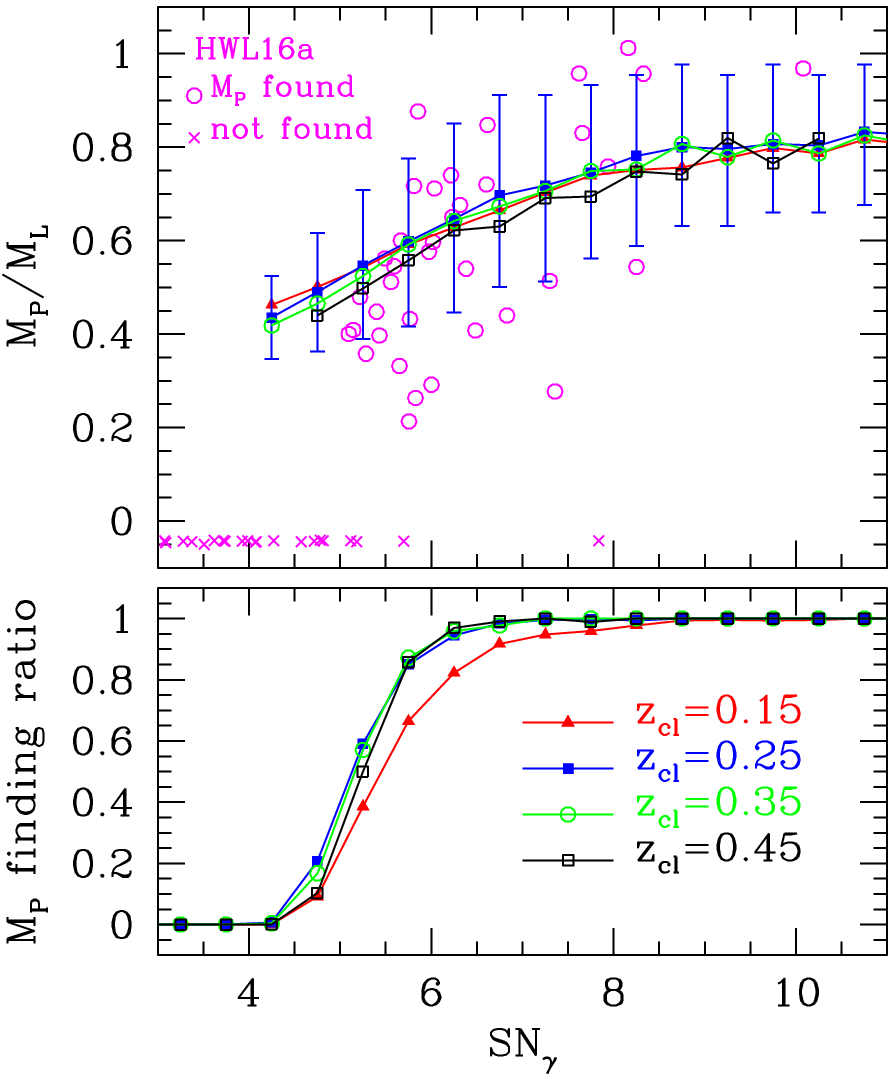}
\end{center}
\caption{{\it Bottom panel:} The $M_P$ finding
  %(the maximum and peak point,
%but not at the low-mass edge of the posterior distribution, of
  %the posterior distribution is found)
  ratios as a function of $SN_\gamma$
evaluated using mock cluster samples are shown.
Different symbols with different colors are for mock cluster samples with different
  redshift $z_{\rm cl}$.
  {\it Top panel:} Statistics of $M_P/M_L$ ($M_{200c}$) are shown as a
  function of $SN_\gamma$. 
  Symbols with error bars (plotted only for $z_{\rm cl}=0.25$, for
  clarity) show the average values and the 16th and
  84th percentiles of mock cluster samples in $SN_\gamma$-bins.
  Magenta symbols show $M_P/M_L$ for individual clusters of HWL16a
  sample \citep{2020PASJ...72...78H}: open circles and crosses (placed
  at $M_P/M_L=-0.05$) are for
  clusters with $M_P$ found, and not found, respectively
  \label{fig:fract_mpost_mlike_snds}}
\end{figure}

An example of the posterior distribution is shown in
Figure~\ref{fig:mock_example} as red contours.
It is seen in the plot that the posterior is truncated at a low-mass side
that is due to the cutoff
imposed on the prior, $Pr=0$ for $SN_\gamma^{\rm NFW}<1$, see equation~(\ref{eq:prior}).
The marginalized one-dimensional posterior of $M$ is truncated at a
low-mass side as well.
Due to this truncation, there are cases in which the maximum {\it and
  peak} point (not at the low-mass edge of the posterior distribution) of
the posterior distribution does not exist, and thus $M_P$ is not determined.
We evaluate ratios of cases that do have $M_P$ for all the cases as a
function of $SN_\gamma$ and present the result in the bottom panel of
Figure~\ref{fig:fract_mpost_mlike_snds}. 
It is seen in the Figure that the ratio drops at $SN_\gamma\sim 6$ and
below, and is zero below $SN_\gamma\sim 4$.
The reason for this is that the smaller $SN_\gamma$ is, the
broader the likelihood function becomes, with its maximum point being
generally at
smaller mass, resulting in its posterior being more affected by the
truncation of the prior.
This makes the limitations of our correction method evident; for our method
to be workable, a high signal-to-noise ratio measurement of
$SN_\gamma \gtsim 5$ is required.

%
%%%%% Fig mtrue_mlike_snpeak_snlens_z0.25 --- fig-8
%
\begin{figure}
\begin{center}
  \includegraphics[height=82mm,angle=270]{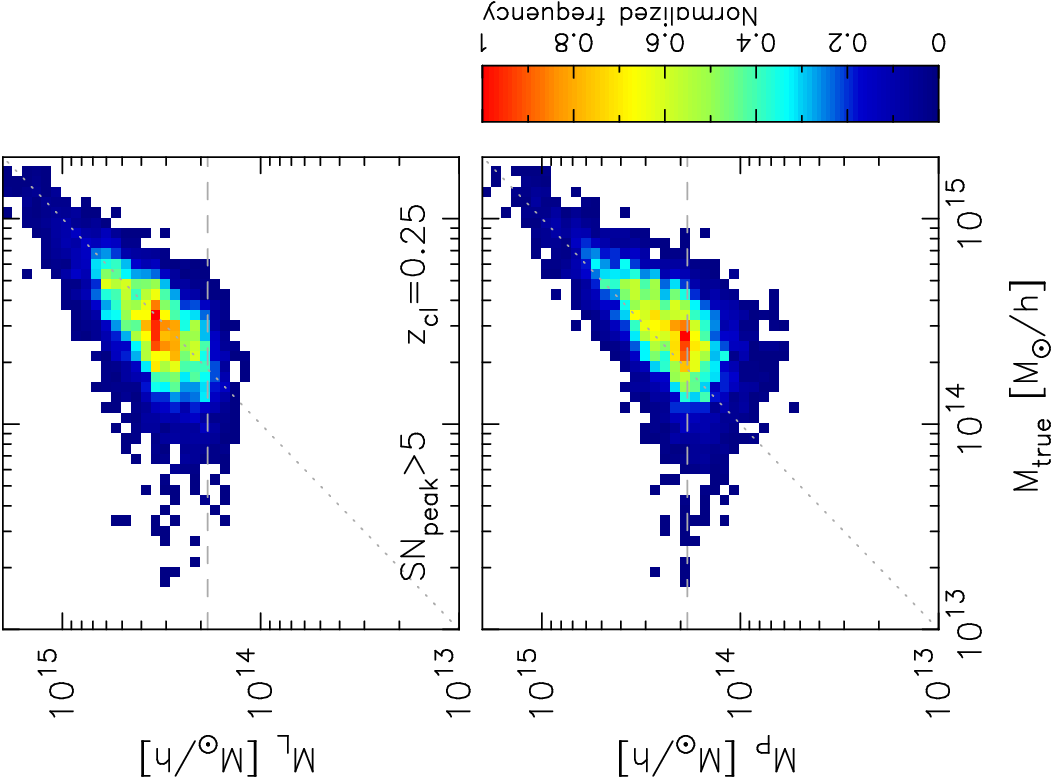}
\end{center}
\caption{Frequency distributions of mock weak lensing clusters selected with
  $SN_{\rm peak}>5$ in $M_{\rm true}$--$M_L$ plane (top panel) and
  $M_{\rm true}$--$M_P$ plane (bottom panel).
  All the cluster masses are in $M_{200c}$.
  Note in both the panels, a sample of mock clusters with $M_P$ found is used.
  The results from the mock sample of $z_{\rm cl}=0.25$ are shown. The
  horizontal dashed line in each panel shows the empirical mass scale
  $M_s$ obtained by the condition, equation~(\ref{eq:Ms}).
  \label{fig:mock_Mtrue-ML-MP}}
\end{figure}

%
%%%%% Fig bias_mlike_snpeak_snlens --- fig-9
%
\begin{figure}
\begin{center}
  \includegraphics[width=82mm]{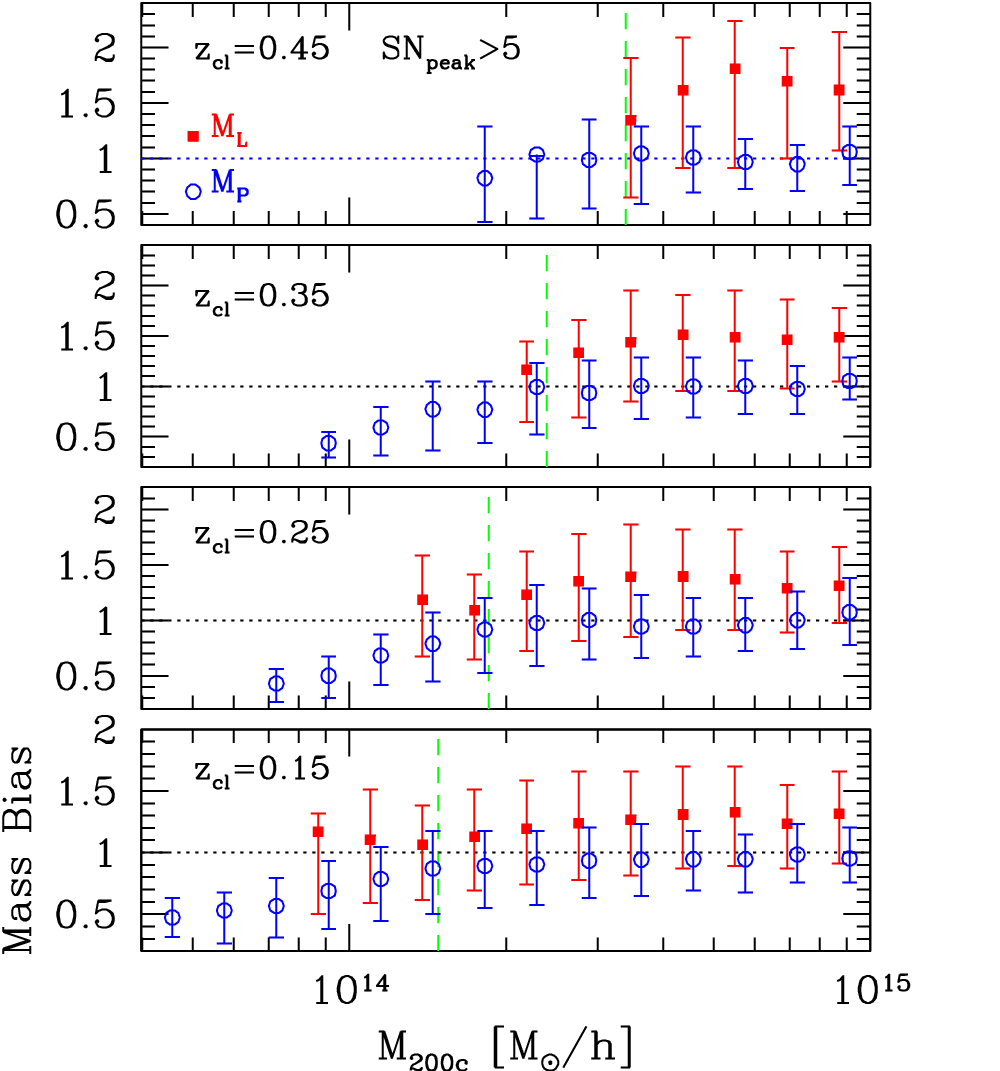}
\end{center}
\caption{Mass biases defined either by $b=M_L/M_{\rm true}$ as a
  function of $M_L$ (red filled squares) or by $b=M_P/M_{\rm true}$ as a
  function of $M_P$ (blue open circles) are shown.
  All the cluster masses are in $M_{200c}$.
  Symbols show the average values of mock cluster samples in $M$-bins, and error
  bars show the 16th and 84th percentiles of the samples in each bin.
  A sample of mock clusters selected with $SN_{\rm peak}>5$ is used.
  Note that the results are slightly shifted in the horizontal direction
  for clarity.
  The vertical dashed line in each panel shows the empirical mass scale
  $M_S$, below which the bias values are affected by the cluster
  selection (see main text in Section~\ref{sec:correction:test} for
  details).  
  \label{fig:bias_mobs_mpost_primc_zall}}
\end{figure}

Next, we examine the performance of our correction method.
The frequency distribution of mock weak lensing clusters selected with
$SN_{\rm peak}>5$ in the $M_{\rm true}$--$M_P$ plane is shown in the bottom
panel of Figure~\ref{fig:mock_Mtrue-ML-MP}, and the distribution of the
same sample but in $M_{\rm true}$--$M_L$ plane is shown in the top panel.
Comparing the two plots, one can see a trend that the distribution moves
to down as a result of the correction.
The mass bias after the correction, defined by $b=M_P/M_{\rm true}$, as a
function of $M_P$ is shown by open circles with error bars in
Figure~\ref{fig:mock_Mtrue-ML-MP}. 
In that plot, a trend that the bias decreases at lower-mass ranges is
seen.
This is a consequence of sample selection with $SN_{\rm peak}$; to be
more specific, only clusters with $M_P < M_{\rm true}$ exist at a
lower-$M_P$ range (see Figure~\ref{fig:mock_Mtrue-ML-MP}).
We empirically define a mass scale (denoted by $M_s$), below which the
averaged bias is affected by the sample selection, by introducing the
following condition, 
\begin{equation}
\label{eq:Ms}
\int dc P(c|M_s) SN_{\rm NFW}(M_s,c)=4,
\end{equation}
where $SN_{\rm NFW}$ is the model prediction of the weak
lensing peak height defined in the same manner as $SN_{\rm peak}$,
equation~(\ref{eq:SNpeak}). 
The derived mass scales are shown in 
Figure~\ref{fig:bias_mobs_mpost_primc_zall} as vertical dashed lines, 
and in Figure~\ref{fig:mock_Mtrue-ML-MP} as horizontal dashed lines.
It is seen from Figure~\ref{fig:bias_mobs_mpost_primc_zall} that 
in the mass range $M_P>M_s$, our correction method works well, with
resulting bin averaged mass biases being close to unity within
$\sim 10$ percent.

Finally, we check if the marginalized one-dimensional posterior
distributions of the mass give proper credible intervals of the true mass.
Since the posterior distribution is truncated at the low-mass side, the
marginalized 1D posterior is truncated as well.
Taking the case shown in Figure~\ref{fig:mock_example} as an example, as can
be seen in the right-hand panel, the conventional 1-$\sigma$ credible
interval\footnote{We
define the 1(2)-$\sigma$ credible interval by the following conventional
manner;
$\Delta\chi^2=-2\log P(M)/P_{max}<1(4)$, where $P(M)$ is the
marginalized 1D posterior distribution of mass, and $P_{max}$ is its
maximum value.}
is safely determined, but the 2-$\sigma$ interval is not.
It is found that in 74(38) percent of all the mock clusters that do have
$M_P$, the 1(2)-$\sigma$ credible interval of 1D
posterior distribution is determined. 
Then, it is found that in 66(95) percent of the cases with the
1(2)-$\sigma$ credible interval being determined, $M_{\rm true}$ is within
the 1(2)-$\sigma$ credible interval of the 1D posterior distribution.
Therefore, we may conclude that if credible intervals are determined,
the marginalized 1D posterior distributions of the mass give proper
credible intervals of the true mass.

%
%%%%%% Sec-5: HWL16a sample %%%%%%%%%%%%%%%%%%%%%%%%%%%%%%%%%%%%%%%%%%%%%%%%%%
%
\section{Application to HWL16a sample}
\label{sec:HWL16a}

We apply our correction method of the excess up-scattering mass bias to
the weak lensing shear-selected cluster sample of 
\citet{2020PASJ...72...78H} (which we call ``HWL16a sample'').
They conducted weak lensing cluster search using Hyper Suprime-Cam
Subaru Strategic Program (HSC survey) first-year data
\citep{2018PASJ...70S...4A,2018PASJ...70S...8A}. 
They searched for high peaks in weak lensing mass maps covering an effective
area of $\sim 120$ deg$^2$ generated using HSC survey first-year
weak lensing shape catalog \citep{2018PASJ...70S..25M}.
They found 124 high peaks with $SN_{\rm peak} > 5$, and cross-matched
them with the public optical cluster catalog 
\citep[CAMIRA,][]{2018PASJ...70S..20O} constructed from the same
HSC survey data to identify cluster counter-parts of the peaks.
They defined the sub-sample of 64 secure clusters, and performed
weak lensing mass estimate for 61 clusters located at $z_{\rm cl}<0.7$ 
among the 64 clusters.

The weak lensing mass estimation method of \citet{2020PASJ...72...78H} is
basically identical with one adopted in our mock weak lensing cluster
analysis described in Section~\ref{sec:analyses} (in fact, our
method is based on theirs) except for their
use of photo-$z$ probability distribution functions to select
background source galaxies and to evaluate $\Sigma_{\rm cr}^{-1}$
\citep[see for details Appendix 3 of ][and reference therein]{2020PASJ...72...78H}. 
Its impact on mass estimate was studies in
\citet{2019ApJ...875...63M,2020ApJ...890..148U}, they found 
that the induced uncertainty in estimated mass is less than 2 percent
which is much smaller than the statistical uncertainty. 
Therefore, we directly apply our correction method,  
equation~(\ref{eq:posterior}), with the prior, equation~(\ref{eq:prior}),
to HWL16a secure clusters with likelihood
functions derived in \citet{2020PASJ...72...78H}.
In so doing, we adopt the cosmological model from Planck 2018 result
\citep{2020A&A...641A...6P}, and we consider two cases, $\Delta=200$ and
$\Delta=500$.

%
%%%%% Table-1
%
%%%%%% Table: cluster mass  %%%%%%%%%%%%%%%%%%%%%%%%%%%%%%%%%%%%%%%%%%%%%%%%%%
%
\begin{longtable}{lcclrcccc}
\caption{Results of the weak lensing mass estimate for HWL16a sample are summarized.
Columns 6--7 are for the over-density parameter relative to the critical density 
of $\Delta =200$, and Columns 8--9 are for $\Delta =500$.
$M_L$ and $M_P$ are the estimated cluster mass before and after the correction for 
the excess up-scattering mass bias is applied, respectively.
Errors are taken from the 1-$\sigma$ confidence interval of marginalized 1D probability 
distributions.
"N/A" means that a corresponding value is not obtained.
\label{table:clustermass}}
\hline
{} & RA & Dec & {} & {} & \multicolumn{2}{c}{$M_{200c}[10^{14}M_\odot h^{-1}]$} & \multicolumn{2}{c}{$M_{500c}[10^{14}M_\odot h^{-1}]$} \\
\cmidrule(lr){6-7}
\cmidrule(lr){8-9}
ID & \multicolumn{2}{c}{J2000.0[$^{\circ}$]} &$z_{\rm cl}$ & $SN_\gamma$ & $M_L$ & $M_P$ & $M_L$ & $M_P$ \\
\hline
\endhead
HWL16a-002 & 30.4273 & $-5.0219$ & 0.234 & 6.62 & $3.35_{-1.35}^{+1.08}$ & $2.84_{-1.05}^{+1.08}$	& $2.51_{-0.93}^{+0.60}$ & $1.93_{-0.74}^{+0.79}$ \\
HWL16a-003 & 31.2073 & $-3.0587$ & 0.553 & 2.72 & $2.57_{-1.51}^{+1.53}$ & N/A	& $1.95_{-1.14}^{+0.86}$ & N/A \\
HWL16a-005 & 31.4584 & $-3.3714$ & 0.167 & 3.51 & $0.99_{-0.35}^{+0.78}$ & N/A	& $0.83_{-0.33}^{+0.53}$ & N/A \\
HWL16a-007 & 33.1112 & $-5.6214$ & 0.287 & 7.36 & $7.69_{-4.22}^{+0.65}$ & $2.13_{-1.24}^{+1.29}$	& $2.98_{-1.20}^{+1.04}$ & $1.48_{-0.87}^{+0.91}$ \\
HWL16a-012 & 35.4434 & $-3.7668$ & 0.430 & 6.02 & $5.39_{-2.64}^{+2.45}$ & $3.22_{-1.45}^{+1.48}$	& $3.71_{-1.63}^{+0.93}$ & $2.22_{-1.03}^{+1.06}$ \\
HWL16a-013 & 36.1229 & $-4.2378$ & 0.264 & 5.56 & $2.62_{-1.37}^{+1.05}$ & $1.34_{-0.98}^{+0.93}$	& $1.80_{-0.86}^{+0.45}$ & $0.87_{-{\rm N/A}}^{+0.68}$ \\
HWL16a-014 & 36.3758 & $-4.2496$ & 0.155 & 5.65 & $2.41_{-1.41}^{+0.98}$ & $0.80_{-{\rm N/A}}^{+0.92}$	& $1.56_{-0.83}^{+0.41}$ & N/A \\
HWL16a-016 & 37.3963 & $-3.6121$ & 0.312 & 6.83 & $6.09_{-2.72}^{+1.73}$ & $2.68_{-1.18}^{+1.23}$	& $3.35_{-1.38}^{+0.75}$ & $1.85_{-0.82}^{+0.86}$ \\
HWL16a-017 & 37.5572 & $-5.6526$ & 0.500 & 3.08 & $4.06_{-3.02}^{+1.27}$ & N/A	& $2.16_{-1.44}^{+0.72}$ & N/A \\
HWL16a-020 & 37.9163 & $-4.8799$ & 0.186 & 10.08 & $5.04_{-1.72}^{+1.44}$ & $4.88_{-1.25}^{+1.26}$	& $3.78_{-1.16}^{+0.84}$ & $3.43_{-0.90}^{+0.89}$ \\
HWL16a-022 & 38.1580 & $-4.7513$ & 0.276 & 5.98 & $2.88_{-1.40}^{+1.10}$ & $1.66_{-1.02}^{+1.04}$	& $2.05_{-0.92}^{+0.52}$ & $1.10_{-0.78}^{+0.76}$ \\
HWL16a-023 & 38.3915 & $-5.5027$ & 0.420 & 4.07 & $1.99_{-0.78}^{+1.08}$ & N/A	& $1.60_{-0.66}^{+0.65}$ & N/A \\
HWL16a-024 & 129.3206 & $1.6069$ & 0.360 & 3.74 & $3.26_{-1.93}^{+1.34}$ & N/A	& $2.19_{-1.21}^{+0.68}$ & N/A \\
HWL16a-026 & 130.5895 & $1.6473$ & 0.424 & 6.48 & $11.77_{-4.51}^{+3.63}$ & $4.80_{-2.01}^{+1.98}$	& $6.54_{-2.35}^{+1.34}$ & $3.33_{-1.37}^{+1.37}$ \\
HWL16a-028 & 133.1296 & $0.4041$ & 0.270 & 6.23 & $3.66_{-1.68}^{+1.44}$ & $2.38_{-1.06}^{+1.10}$	& $2.54_{-1.08}^{+0.68}$ & $1.61_{-0.75}^{+0.79}$ \\
HWL16a-032 & 138.4612 & $-0.7631$ & 0.285 & 5.67 & $3.58_{-1.70}^{+1.36}$ & $2.15_{-1.07}^{+1.11}$	& $2.46_{-1.08}^{+0.65}$ & $1.44_{-0.76}^{+0.80}$ \\
HWL16a-034 & 139.0387 & $-0.3966$ & 0.315 & 7.94 & $9.16_{-2.77}^{+2.71}$ & $6.95_{-1.91}^{+1.86}$	& $6.32_{-1.74}^{+1.37}$ & $4.85_{-1.32}^{+1.30}$ \\
HWL16a-035 & 139.3198 & $0.9985$ & 0.344 & 3.93 & $2.93_{-2.02}^{+1.05}$ & N/A	& $1.28_{-0.79}^{+0.75}$ & N/A \\
HWL16a-036 & 139.3405 & $3.8281$ & 0.420 & 5.22 & $3.94_{-1.93}^{+1.58}$ & $1.89_{-{\rm N/A}}^{+1.38}$	& $2.75_{-1.27}^{+0.81}$ & $1.23_{-{\rm N/A}}^{+1.01}$ \\
HWL16a-037 & 140.0954 & $1.5748$ & 0.697 & 2.98 & $13.48_{-10.51}^{+3.97}$ & N/A	& $6.98_{-5.01}^{+1.67}$ & N/A \\
HWL16a-038 & 140.1431 & $0.7907$ & 0.463 & 3.09 & $1.51_{-0.64}^{+1.67}$ & N/A	& $1.27_{-0.59}^{+1.04}$ & N/A \\
HWL16a-039 & 140.4154 & $-0.2491$ & 0.310 & 5.18 & $2.52_{-1.53}^{+1.16}$ & N/A	& $1.70_{-0.96}^{+0.46}$ & N/A \\
HWL16a-041 & 140.6790 & $2.1327$ & 0.194 & 5.28 & $2.82_{-1.65}^{+1.17}$ & $1.01_{-{\rm N/A}}^{+1.04}$	& $1.85_{-0.97}^{+0.43}$ & $0.58_{-{\rm N/A}}^{+0.81}$ \\
HWL16a-045 & 177.2946 & $0.3636$ & 0.260 & 4.08 & $2.40_{-1.82}^{+0.27}$ & N/A	& $0.99_{-0.57}^{+0.47}$ & N/A \\
HWL16a-046 & 177.5842 & $-0.6009$ & 0.135 & 6.22 & $2.00_{-0.83}^{+0.80}$ & $1.48_{-0.79}^{+0.86}$	& $1.53_{-0.61}^{+0.45}$ & $0.96_{-0.59}^{+0.64}$ \\
HWL16a-047 & 178.0615 & $0.5187$ & 0.472 & 5.81 & $4.03_{-0.87}^{+1.70}$ & $2.89_{-{\rm N/A}}^{+1.91}$	& $3.38_{-0.84}^{+1.09}$ & $1.88_{-{\rm N/A}}^{+1.41}$ \\
HWL16a-050 & 178.8288 & $0.8712$ & 0.481 & 3.73 & $4.48_{-3.19}^{+1.35}$ & N/A	& $1.76_{-1.21}^{+1.02}$ & N/A \\
HWL16a-051 & 179.0517 & $-0.3490$ & 0.254 & 8.25 & $7.10_{-2.63}^{+2.10}$ & $3.86_{-1.33}^{+1.37}$	& $4.07_{-1.37}^{+0.88}$ & $2.63_{-0.90}^{+0.94}$ \\
HWL16a-052 & 179.6138 & $-0.0412$ & 0.252 & 4.72 & $2.91_{-2.06}^{+0.53}$ & N/A	& $1.41_{-0.80}^{+0.44}$ & N/A \\
HWL16a-053 & 180.4286 & $-0.1839$ & 0.167 & 7.30 & $3.68_{-2.07}^{+1.70}$ & $1.89_{-1.00}^{+1.09}$	& $2.44_{-1.23}^{+0.53}$ & $1.29_{-0.73}^{+0.78}$ \\
HWL16a-056 & 181.3878 & $-0.6432$ & 0.470 & 4.81 & $4.82_{-2.47}^{+1.83}$ & N/A	& $2.98_{-1.42}^{+0.95}$ & N/A \\
HWL16a-057 & 210.7874 & $-0.3084$ & 0.450 & 4.27 & $5.79_{-3.28}^{+2.19}$ & N/A	& $3.40_{-1.80}^{+1.00}$ & N/A \\
HWL16a-058 & 211.2955 & $-0.1472$ & 0.248 & 5.10 & $3.07_{-1.62}^{+1.10}$ & $1.23_{-{\rm N/A}}^{+1.13}$	& $2.02_{-0.96}^{+0.54}$ & $0.77_{-{\rm N/A}}^{+0.84}$ \\
HWL16a-059 & 211.7872 & $-0.2717$ & 0.561 & 3.28 & $2.83_{-2.05}^{+1.53}$ & N/A	& $1.95_{-1.37}^{+1.02}$ & N/A \\
HWL16a-060 & 211.9925 & $-0.4857$ & 0.469 & 5.83 & $8.97_{-4.11}^{+2.57}$ & $2.36_{-{\rm N/A}}^{+2.05}$	& $4.70_{-1.90}^{+1.35}$ & $1.63_{-{\rm N/A}}^{+1.44}$ \\
HWL16a-064 & 213.7770 & $-0.4892$ & 0.144 & 7.84 & $5.65_{-3.99}^{+0.10}$ & N/A	& $1.87_{-1.02}^{+0.76}$ & N/A \\
HWL16a-070 & 215.2574 & $0.3665$ & 0.645 & 4.78 & $10.58_{-5.63}^{+3.77}$ & N/A	& $5.54_{-2.76}^{+1.67}$ & N/A \\
HWL16a-071 & 215.9165 & $0.4491$ & 0.534 & 3.99 & $5.25_{-3.12}^{+1.82}$ & N/A	& $2.97_{-1.56}^{+0.95}$ & N/A \\
HWL16a-076 & 216.7785 & $0.7267$ & 0.296 & 8.33 & $3.26_{-1.39}^{+1.38}$ & $3.12_{-0.97}^{+1.00}$	& $2.51_{-0.98}^{+0.73}$ & $2.18_{-0.70}^{+0.72}$ \\
HWL16a-077 & 216.8310 & $0.9541$ & 0.294 & 5.49 & $2.42_{-1.07}^{+0.87}$ & $1.36_{-{\rm N/A}}^{+1.03}$	& $1.80_{-0.76}^{+0.48}$ & $0.87_{-{\rm N/A}}^{+0.74}$ \\
HWL16a-080 & 217.6808 & $0.8093$ & 0.312 & 6.60 & $5.04_{-2.19}^{+1.92}$ & $3.63_{-1.32}^{+1.35}$	& $3.55_{-1.37}^{+0.86}$ & $2.50_{-0.93}^{+0.97}$ \\
HWL16a-081 & 218.8457 & $-0.3931$ & 0.283 & 6.03 & $1.49_{-0.28}^{+0.77}$ & $1.06_{-{\rm N/A}}^{+0.91}$	& $1.24_{-0.27}^{+0.56}$ & $0.57_{-{\rm N/A}}^{+0.72}$ \\
HWL16a-084 & 220.0846 & $-0.6101$ & 0.549 & 3.62 & $6.35_{-4.62}^{+1.30}$ & N/A	& $2.56_{-1.62}^{+1.24}$ & N/A \\
HWL16a-088 & 220.7952 & $1.0452$ & 0.528 & 3.72 & $4.97_{-2.97}^{+2.52}$ & N/A	& $3.49_{-1.97}^{+1.11}$ & N/A \\
HWL16a-090 & 221.1442 & $0.2464$ & 0.295 & 5.14 & $3.18_{-1.77}^{+1.22}$ & $1.30_{-{\rm N/A}}^{+1.11}$	& $2.09_{-1.04}^{+0.50}$ & $0.85_{-{\rm N/A}}^{+0.81}$ \\
HWL16a-091 & 221.1917 & $-0.6694$ & 0.523 & 5.77 & $9.83_{-4.50}^{+3.86}$ & $4.25_{-2.00}^{+2.00}$	& $6.01_{-2.54}^{+1.29}$ & $2.93_{-1.40}^{+1.44}$ \\
HWL16a-093 & 221.3335 & $0.1116$ & 0.286 & 6.38 & $3.74_{-1.70}^{+1.30}$ & $2.02_{-1.00}^{+1.05}$	& $2.39_{-1.02}^{+0.62}$ & $1.36_{-0.70}^{+0.74}$ \\
HWL16a-094 & 223.0801 & $0.1689$ & 0.592 & 5.76 & $25.86_{-11.72}^{+N/A}$ & $5.52_{-3.88}^{+3.07}$	& $13.70_{-6.07}^{+1.42}$ & $3.89_{-2.68}^{+2.27}$ \\
HWL16a-095 & 223.0929 & $-0.9713$ & 0.304 & 7.66 & $5.42_{-1.89}^{+1.68}$ & $4.50_{-1.33}^{+1.33}$	& $3.89_{-1.24}^{+0.92}$ & $3.13_{-0.94}^{+0.94}$ \\
HWL16a-097 & 224.2746 & $0.1164$ & 0.220 & 4.57 & $2.18_{-1.33}^{+0.78}$ & N/A	& $1.45_{-0.81}^{+0.37}$ & N/A \\
HWL16a-098 & 224.6567 & $0.4858$ & 0.395 & 5.12 & $5.17_{-3.35}^{+0.93}$ & N/A	& $2.46_{-1.29}^{+0.67}$ & N/A \\
HWL16a-101 & 245.3758 & $42.7648$ & 0.152 & 8.16 & $3.23_{-1.06}^{+0.94}$ & $3.27_{-1.04}^{+1.08}$	& $2.54_{-0.80}^{+0.58}$ & $2.26_{-0.76}^{+0.79}$ \\
HWL16a-102 & 246.1339 & $43.3203$ & 0.287 & 5.70 & $3.61_{-1.94}^{+1.24}$ & N/A	& $1.93_{-0.98}^{+0.63}$ & N/A \\
HWL16a-103 & 246.5173 & $43.7147$ & 0.260 & 5.59 & $1.98_{-0.78}^{+0.71}$ & $1.08_{-{\rm N/A}}^{+0.94}$	& $1.52_{-0.60}^{+0.45}$ & $0.61_{-{\rm N/A}}^{+0.71}$ \\
HWL16a-104 & 333.0522 & $-0.1334$ & 0.350 & 6.00 & $6.79_{-3.26}^{+1.49}$ & $1.98_{-1.32}^{+1.28}$	& $3.20_{-1.32}^{+0.86}$ & $1.36_{-0.95}^{+0.89}$ \\
HWL16a-107 & 333.5929 & $0.7956$ & 0.308 & 5.40 & $3.73_{-1.82}^{+1.38}$ & $1.67_{-1.17}^{+1.11}$	& $2.33_{-1.07}^{+0.66}$ & $1.11_{-{\rm N/A}}^{+0.79}$ \\
HWL16a-110 & 335.2140 & $0.9704$ & 0.323 & 5.43 & $5.06_{-2.26}^{+1.78}$ & $2.01_{-{\rm N/A}}^{+1.34}$	& $2.98_{-1.26}^{+0.87}$ & $1.36_{-{\rm N/A}}^{+0.95}$ \\
HWL16a-112 & 336.0366 & $0.3331$ & 0.154 & 7.62 & $3.30_{-1.16}^{+0.99}$ & $3.16_{-1.05}^{+1.09}$	& $2.56_{-0.85}^{+0.59}$ & $2.17_{-0.76}^{+0.80}$ \\
HWL16a-114 & 336.4066 & $-0.3068$ & 0.402 & 3.37 & $2.30_{-1.36}^{+1.07}$ & N/A	& $1.63_{-0.92}^{+0.52}$ & N/A \\
HWL16a-115 & 336.4217 & $1.0730$ & 0.281 & 5.86 & $2.59_{-0.76}^{+0.99}$ & $2.27_{-1.05}^{+1.11}$	& $2.10_{-0.64}^{+0.61}$ & $1.52_{-0.76}^{+0.81}$ \\
HWL16a-117 & 337.1293 & $1.7135$ & 0.338 & 6.32 & $5.87_{-2.50}^{+2.26}$ & $3.97_{-1.49}^{+1.52}$	& $4.06_{-1.56}^{+1.01}$ & $2.75_{-1.05}^{+1.08}$ \\
\hline
\end{longtable}

%
%%%%% Fig bias_mlike_snpeak_snlens --- fig-10
%
\begin{figure}
\begin{center}
  \includegraphics[width=82mm]{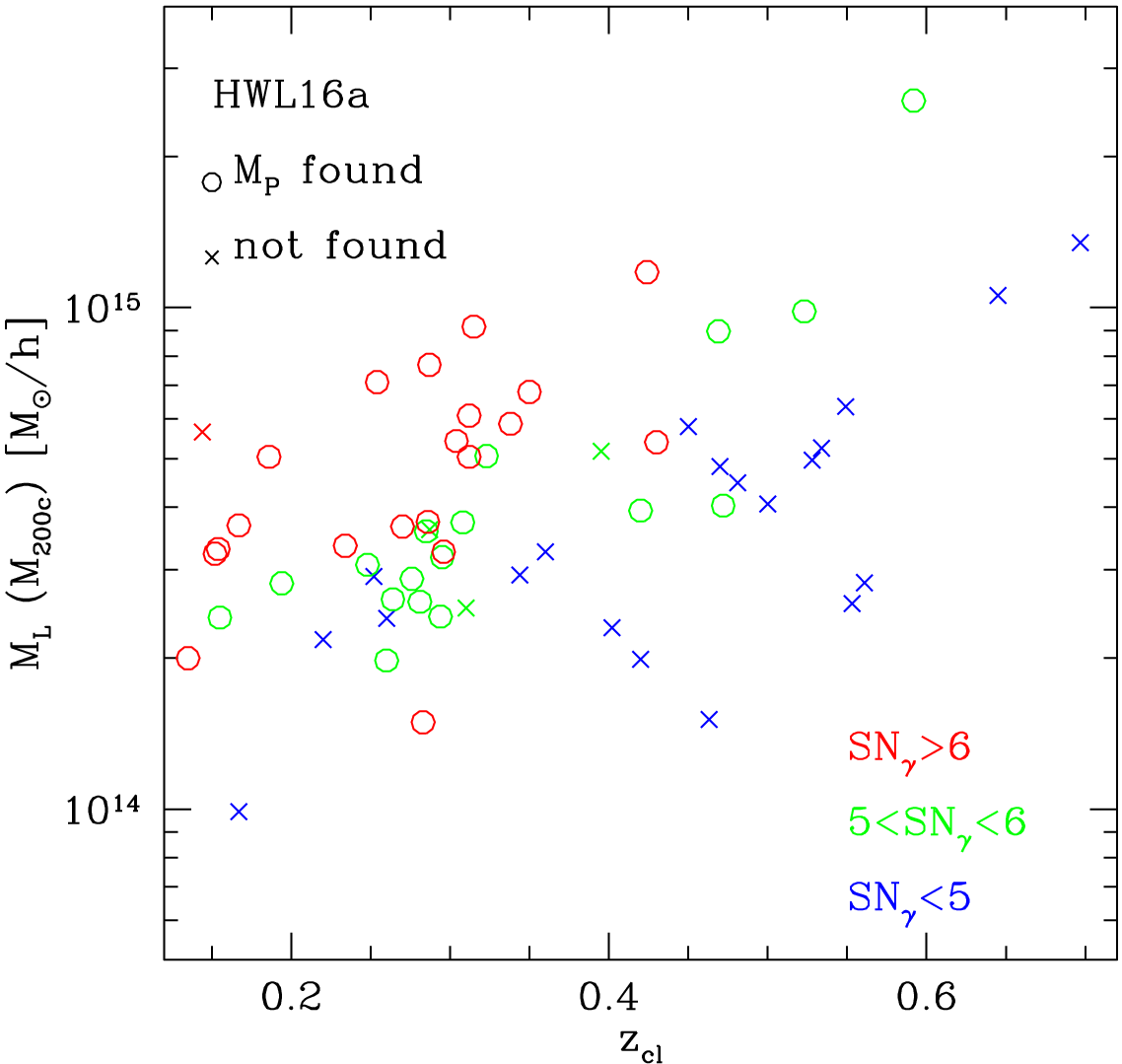}
\end{center}
\caption{Distribution of the HWL16a weak lensing cluster sample in
  $M_L$--$z_{\rm cl}$ plane. The weak lensing masses, $M_L$, are
  re-derived adopting the maximum point of the likelihood function,
  ${\cal{L}}(M,c)$, as the best-fitting model (in the original study by
  \citet{2020PASJ...72...78H}, the peak of the marginalized 1D
  likelihood function was adopted as a point estimator of the weak
  lensing mass). The open circles are clusters with $M_P$ being found,
  whereas the crosses are ones with $M_P$ being not found. Different
  colors are for different ranges of cluster's $SN_\gamma$ values, see
  colored labels in the plot. 
  \label{fig:mlike_zcl_200c_mp_pl16}}
\end{figure}

The derived cluster
masses\footnote{
%%%% footnote
The weak lensing masses, $M_L$, derived in this work differ from those
presented in \citet{2020PASJ...72...78H} due to the different
choices for the point estimator of the best-fitting mass.
In \citet{2020PASJ...72...78H},
peaks of marginalized 1D probability distributions of the mass was taken
as the best-fitting mass.
They performed the marginalization over a range of $0.01<c<30$
with a uniform prior for $c$, which tends to result in a
smaller best-fitting mass, because of the anti-correlation between the mass
and concentration parameter with a poorly constrained 2D distribution,
especially in a larger-$c$ (thus a smaller-$M$) region (see Figure
\ref{fig:mock_example} for an example).
Since we take the maximum point of 2D likelihood function as the
best-fitting $M_L$, our $M_L$ values are systematically larger than those given in
\citet{2020PASJ...72...78H}.}
%%%% footnote
are summarized in Table~\ref{table:clustermass}.
We also evaluate the signal-to-noise ratio of the measured
$\Delta\Sigma(R)$ profile with the corresponding shape noise
covariance, and we summarize the obtained values in
Table~\ref{table:clustermass}.
Note that the resulting $SN$ is equivalent to $SN_\gamma$ (defined in
equation~(\ref{eq:sngamma})) evaluated for
mock clusters because the tangential shear and $\Delta\Sigma(R)$
are related by equation~(\ref{eq:gammat}), and thus we denote it by 
$SN_\gamma$.

The distribution of HWL16a clusters in the $z_{\rm cl}$--$M_L(M_{}200c)$
plane is shown in
Figure~\ref{fig:mlike_zcl_200c_mp_pl16}.
$M_P$ is found for 37(36) out of the 61 clusters for
$\Delta=200(500)$.
In Figure~\ref{fig:mlike_zcl_200c_mp_pl16}, 
clusters with $M_P$ being found are shown by open circles, whereas ones
without $M_P$ are shown by crosses, and different colors are for
different ranges of cluster’s $SN_\gamma$ values.
It is seen in the figure that clusters with lower-$SN_\gamma$, which are
generally located at a high-$z_{\rm cl}$ and low-$M_L$ region, fail to
obtain $M_P$.
This tendency is also seen in the top panel of
Figure~\ref{fig:fract_mpost_mlike_snds}.
In that figure the distribution in the $SN_\gamma$--$M_P/M_L$ ($M_{200c}$)
plane is shown, where clusters without $M_P$ are placed at $M_P/M_L=-0.05$. 
It is found from this figure that the results from the HWL16a sample
are broadly consistent with results from mock clusters, especially in
the following
two points:
(1) For the HWL16a sample, the $M_P$ finding ratio drops at below 
$SN_\gamma \sim 6$, and no 
HWL16a cluster with $M_P$ successfully found exists at $SN_\gamma < 5$. This
trend is very similar to that of mock clusters. 
(2) The overall trend of $M_P/M_L$ as a function of $SN_\gamma$ for
individual HWL16a cluster is in line with the trend seen in mock clusters.
Finally, we evaluate the mean value of the $M_P/M_L$ ratios among HWL16a
clusters that do have $M_P$, and we obtain $\langle M_P/M_L \rangle = 0.58$
with ${\rm RMS}=0.21$ (for $M_{200c}$).

%
%%%%%% Sec6: Summary and discussions %%%%%%%%%%%%%%%%%%%%%%%%%%%%%%%%%%%%%%%%%%%%%%%%%%
%
\section{Summary and discussions}
\label{sec:summary}

We have examined the excess up-scattering mass bias on the weak lensing
mass estimate for shear-selected cluster samples, and have present the
empirical method for mitigating it.

We have generated samples of realistic mock weak lensing clusters for four
redshifts ($z_{\rm cl}=0.15$, 0.25, 0.35 and 0.45) taking into
account the following three components (see
Section~\ref{sec:mock:recipe}); the cluster lensing including deviations
from the NFW profile, lensing contributions from large-scale structures,
and the noise component from intrinsic galaxy shapes. 
We have performed the standard weak lensing mass estimate for the
shear-selected ($SN_{\rm peak}>5$) mock clusters, and found the mass-bin
averaged mass bias is in the range of 
$1.1-1.8$ depending on cluster mass and the redshift.
Another important finding from the mock weak lensing analyses
is that even before the correction for the excess up-scattering
mass bias is applied, the marginalized one-dimensional probability
distributions of the mass for individual clusters give statistically
correct confidence intervals. 
We thus emphasize that the standard weak lensing mass estimation method
itself is working correctly, but the probability of up- and
down-scattering is not symmetric, resulting in the biased best-fitting
masses, on average. 

Our correction method uses the framework of the standard Bayesian 
statistics with the likelihood function from weak lensing mass
estimate based on the standard $\chi^2$ analysis, and the prior of the
probability distribution of the cluster mass and concentration parameter
from recent empirical models (see Section~\ref{sec:correction:method}).
We tested our correction method with mock weak lensing clusters.
It was found that our correction method works well, with resulting
mass-bin averaged mass biases being close to unity within $\sim 10$
percent.

However, our method has limitations that a high signal-to-noise ratio
measurement of $SN_\gamma \gtsim 5$ is required for the method to work
properly. 
This is due to the artificial truncation of the prior on a low-mass
side, which is imposed to avoid a non-physical upturn of the posterior
at a very low-mass range.
A possible cause of the non-physical upturn is a combination of the 
following two factors:
One is the fact that the likelihood function changes only slowly on
the lower-mass side, because the model shear profile approaches
$\gamma_t(\theta)\rightarrow 0$ as the mass decreases.
The other is the power-law shape of halo mass function on a lower-mass
range with a negative power-law index.
One possible way to resolve the non-physical upturn is to modify the prior
so that it includes a condition that clusters are shear-selected;
that is, the probability that a cluster appears as a high peak in a weak
lensing mass map decreases as the cluster mass decreases.
We, however, leave such modifications of the method for a future study.

We applied the method to the HWL16a weak lensing shear-selected cluster
sample, and derived bias corrected weak lensing cluster masses
(summarized in Table~\ref{table:clustermass}).
Because of the limitations of our method, the bias corrected masses, $M_P$,
were obtained only for clusters with a high signal-to-noise ratio of
$SN_\gamma \gtsim 5$. 
In 61 HWL16a clusters, $M_P$ is found for 37(36) clusters for
$\Delta = 200(500)$, among which we found the mean value of the
$M_P/M_L$ to be $\langle M_P/M_L \rangle = 0.58$
with ${\rm RMS}=0.21$ (for $M_{200c}$).

Before closing this paper, we comment on a proper use of the bias
corrected mass, $M_P$:
As is demonstrated in Figure~\ref{fig:demo_exupscatbias}, the excess
up-scattering mass bias arises when one considers the probability of
$M_{\rm true}$ for a given $M_L$, $P(M_{\rm true}|M_L)$, but does not
arise when one considers $P(M_L|M_{\rm true})$ 
(here we assume the error in a weak lensing mass estimate is symmetric
and there are no other systematic or selection effects on a
cluster sample).
It is also important to note that $P(M_P|M_{\rm true})$ is a biased
distribution in the sense that it gives
$\langle M_P\rangle = \int dM_P M_P P(M_P|M_{\rm true})<M_{\rm true}$.
One simple example of a proper use of $M_P$ is to suppose one has a
shear-selected cluster sample, and tries to divide clusters into some
sub-samples so that their mean masses equal to true masses.
In this case, one should make sub-samples with $M_P$, in other words,
one should divide clusters into $M_P$--bins.
Then an expectation value of $M_{\rm true}$ of a sub-samples in a narrow
$M_P$--bins is $M_{\rm true}\simeq M_P$.
The other simple example in which one should not use $M_P$ is if one has
a cluster sample selected 
by a good tracer of the true mass (denoted by $X$), and estimates a mean 
true mass of a sub-sample of clusters in a narrow $X$--bin. 
In this case, one should not use $M_P$ but use $M_L$, because 
$P(M_L|X) \sim P(M_L|M_{\rm true})$ is the unbiased probability distribution which gives 
$\langle M_L \rangle \simeq M_{\rm true}$
(see a discussion in the last paragraph of section~\ref{sec:mock:results}).
We note that in actual recent studies on deriving cluster scaling
relations, for example one between X-ray temperatures and cluster
masses, advanced methods such as Bayesian hierarchical models and
Bayesian population modeling are adopted, in which scatters in measured
quantities, including weak lensing mass, and the selection functions are
statistically accounted for 
\citep{2019ApJ...875...63M,2020MNRAS.492.4528S,2020ApJ...890..148U,2022PASJ...74..175A,2022A&A...661A..11C}.
Therefore, in such a study, one should not use $M_P$ to avoid
double-counting of the excess up-scattering effect.

%
%%%%%% Acknowledgments %%%%%%%%%%%%%%%%%%%%%%%%%%%%%%%%%%%%%%%%%%%%%%%%%%
%
\begin{ack}
We would like to thank the anonymous referee for constructive comments
on the earlier manuscript which improved the paper.
We thank M.~Oguri and K.~Umetsu for useful discussions and comments.
We would like to thank R.~Takahashi for
making full-sky gravitational lensing simulation data publicity
available. 
The dark matter halo mass function and the mass-concentration relation
used in this work were computed by routines wrapped up in {\tt Colossus}
\citep{2018ApJS..239...35D}, we would like to thank B.~Diemer for making
it publicly available.
We would like to thank Nick Kaiser for making the software
{\tt  imcat} publicly available, we have heavily used it in this work.
We would like to thank HSC data analysis software team for their effort
to develop data processing software suite, and HSC data archive team for
their effort to build and to maintain the HSC data archive system.

This work was supported in part by JSPS KAKENHI
Grant Number JP22K03655.

Data analysis were in part carried out on the analysis servers at Center
for Computational Astrophysics, National Astronomical Observatory of
Japan.
Numerical computations were in part carried out on Cray XC30 and
XC50 at Center for Computational Astrophysics, National Astronomical
Observatory of Japan, and also on Cray XC40 at YITP in Kyoto
University.

This paper is partly based on data collected at the Subaru Telescope and
retrieved from the HSC data archive system, which is operated by Subaru
Telescope and Astronomy Data Center (ADC) at NAOJ. Data analysis was in
part carried out with the cooperation of Center for Computational
Astrophysics (CfCA) at NAOJ.  We are honored and grateful for the
opportunity of observing the Universe from Maunakea, which has the
cultural, historical and natural significance in Hawaii. 
\end{ack}

%%%%%%%%%%%%%%%%%%%%%%%%%%%%%%%%%%%%%%%%%%%%%%%%%%%%%%%%%%%%%%%%%%%%%
%%% References
%%%%%%%%%%%%%%%%%%%%%%%%%%%%%%%%%%%%%%%%%%%%%%%%%%%%%%%%%%%%%%%%%%%%%
\bibliographystyle{apj}

\end{document}